\documentclass[twocolumn, tighten, times, usenatbib, useAMS, usenames, dvipsnames]{aastex63}
\usepackage{amsmath}
\usepackage{colortbl}
\usepackage{enumitem}
\hypersetup{linkcolor=magenta,filecolor=cyan}%,urlcolor=magenta,citecolor=blue

\shorttitle{Resolved ISM Properties in TNG50}
\shortauthors{Motwani \& the SMAUG collaboration}

\defcitealias{sh03}{SH03}

\begin{document}

\title{First results from SMAUG: Insights into star formation conditions from spatially-resolved ISM properties in TNG50}

% \correspondingauthor{Bhawna Motwani}
\email{bm2900@columbia.edu}

\author[0000-0002-0045-5684]{Bhawna Motwani}
\affiliation{Department of Astronomy, Columbia University, 550 West 120th Street, New York, NY 10027, USA}
\affiliation{Center for Computational Astrophysics, Flatiron Institute, 162 Fifth Avenue, New York, NY 10010, USA}

\author[0000-0002-3185-1540]{Shy Genel}
\affiliation{Center for Computational Astrophysics, Flatiron Institute, 162 Fifth Avenue, New York, NY 10010, USA}
\affiliation{Columbia Astrophysics Laboratory, Columbia University, 550 West 120th Street, New York, NY 10027, USA}

\author[0000-0003-2630-9228]{Greg L. Bryan}
\affiliation{Department of Astronomy, Columbia University, 550 West 120th Street, New York, NY 10027, USA}
\affiliation{Center for Computational Astrophysics, Flatiron Institute, 162 Fifth Avenue, New York, NY 10010, USA}

\author[0000-0003-2896-3725]{Chang-Goo Kim}
\affiliation{Department of Astrophysical Sciences, Princeton University, Princeton, NJ 08544, USA}
\affiliation{Center for Computational Astrophysics, Flatiron Institute, 162 Fifth Avenue, New York, NY 10010, USA}

\author[0000-0002-2499-9205]{Viraj Pandya}
\affiliation{UCO/Lick Observatory, Department of Astronomy and Astrophysics, University of California, Santa Cruz, CA 95064, USA}
\affiliation{Center for Computational Astrophysics, Flatiron Institute, 162 Fifth Avenue, New York, NY 10010, USA}

\author[0000-0003-2835-8533]{Rachel S. Somerville}
\affiliation{Center for Computational Astrophysics, Flatiron Institute, 162 Fifth Avenue, New York, NY 10010, USA}
\affiliation{Department of Physics and Astronomy, Rutgers University, 136 Frelinghuysen Road, Piscataway, NJ 08854, USA}

\author[0000-0002-9849-877X]{Matthew C. Smith}
\affiliation{Harvard-Smithsonian Center for Astrophysics, 60 Garden Street, Cambridge, MA 02138, USA}

\author[0000-0002-0509-9113]{Eve C. Ostriker}
\affiliation{Department of Astrophysical Sciences, Princeton University, Princeton, NJ 08544, USA}

\author[0000-0001-8421-5890]{Dylan Nelson}
\affiliation{Max-Planck-Institut f\"ur Astrophysik, Karl-Schwarzschild-Str. 1, 85741 Garching, Germany}

\author[0000-0003-1065-9274]{Annalisa Pillepich}
\affiliation{Max-Planck-Institut f\"ur Astronomie, K\"onigstuhl 17, 69117 Heidelberg, Germany}

\author[0000-0002-1975-4449]{John C. Forbes}
\affiliation{Center for Computational Astrophysics, Flatiron Institute, 162 Fifth Avenue, New York, NY 10010, USA}

\author[0000-0002-2545-5752]{Francesco Belfiore}
\affiliation{European Southern Observatory, Karl-Schwarzschild-Strasse 2, D-85748 Garching bei M\"unchen, Germany}
\affiliation{INAF – Osservatorio Astrofisico di Arcetri, Largo E. Fermi 5, I-50157 Firenze, Italy}

\author[0000-0003-3308-2420]{R\"udiger Pakmor}
\affiliation{Heidelberg Institute for Theoretical Studies, Schlo{\ss}-Wolfsbrunnenweg 35, 69118 Heidelberg, Germany}

\author[0000-0001-6950-1629]{Lars Hernquist}
\affiliation{Harvard-Smithsonian Center for Astrophysics, 60 Garden Street, Cambridge, MA 02138, USA}

\begin{abstract}
Physical and chemical properties of the interstellar medium (ISM) at sub-galactic ($\sim$kpc) scales play an indispensable role in controlling the ability of gas to form stars. As part of the SMAUG (Simulating Multiscale Astrophysics to Understand Galaxies) project, in this paper, we use the TNG50 cosmological simulation to explore the physical parameter space of 8 resolved ISM properties in star-forming regions to constrain the areas of this hyperspace over which most star-forming environments exist. We deconstruct our simulated galaxies spanning a wide range of mass (M$_\star = 10^{7-11}$ M$_\odot$) and redshift ($0 \leq z \leq 3$) into kpc-sized regions, and statistically analyze the gas/stellar surface densities, gas metallicity, vertical stellar velocity dispersion, epicyclic frequency and dark-matter volumetric density representative of each region in the context of their star formation activity and galactic environment (radial galactocentric location). By examining the star formation rate (SFR) weighted distributions of these properties, we show that stars primarily form in two spatially distinct environmental regimes, which are brought about by an underlying bi-component radial SFR surface density profile in galaxies. We examine how the relative prominence of these two regimes depends on host galaxy mass and cosmic time. We also compare our findings with those from integral field spectroscopy observations and achieve a good overall agreement. Further, using dimensionality reduction, we characterise the aforementioned hyperspace to reveal a high-degree of multicollinearity in relationships amongst ISM properties that drive the distribution of star formation at kpc-scales. Based on this, we show that a reduced 3D representation underpinned by a multi-variate radius relationship is sufficient to capture most of the variance in the original 8D space.

\end{abstract}


\section{Introduction}
Local characteristics of the ISM drive a complex interplay of processes at the sub-galactic scale which regulate the star formation and feedback in a galaxy \citep[e.g.,][]{leroy08,leroy13,shi18,dey19,sun20}. Gravitational instabilities on kiloparsec (kpc) scales heavily influence the lifecycle and properties of giant molecular clouds, thereby setting up star formation and controlling its efficiency in different regions of the galaxy \citep[e.g.,][]{elmegreen87,elmegreen91,kim01,kim02,kim06,dobbs08,dobbs11,bournaud10,renaud13}. Feedback on these scales influences the dynamical state of the gas by limiting its evolution toward high densities and dispersing the clouds \citep[e.g.,][]{hopkins12a,kim13,kim15b,semenov17,semenov18}. Additionally, hydrodynamical interaction between the hot and cold ISM on kpc-scales facilitates the acceleration of galactic outflows \citep[e.g.,][]{hopkins12b,muratov15,angles-alcazar17,kimostriker18} that subsequently suppress star formation through the depletion of cold gas, and prevention of future gas cooling and accretion \citep[e.g.,][]{bouche10, dave12, lilly13,forbes14b,rodriguezpuebla16}. As such, by virtue of controlling the incidence and effects of star formation, physical properties of the interstellar medium on kpc-scales play a vital role in modulating the overall baryon cycle of galaxies \citep{naabostriker17, someville15}. 

Deciphering the link between star formation and galactic structure is a multi-scale problem, ranging from $\sim$100 pc scale of molecular cloud collapse to the $10^{5-6}$ pc scale of the circumgalactic medium. Due to the steep computational challenge of simulating the vast range of scales involved, modern simulations of galaxy formation must implement only approximate \emph{sub-grid} treatments of small-scale physical processes, smoothing over much of the complexity at or below cloud-scales \citep{naabostriker17}. Consequently, while results from contemporary large-scale cosmological simulations have been shown to match the integrated stellar mass abundances and star formation rates (SFRs) of observed galaxies \citep[see][for a review]{someville15}, they invoke simplified treatments of the underlying, often less-understood, small-scale processes such as star formation and ISM physics, making the treatment less realistic. Understanding the conditions that govern the onset of star formation, and their interrelationships, is therefore, of fundamental importance to the modern theory of galaxy formation and evolution \citep{morselli18, trayford18, orr18, chruslinska19}. The conditions in which stars form is a crucial ingredient that must be firmly constrained in order to improve our models of star formation and feedback. Beyond star formation, characterising the physical properties of galaxies in a spatially-resolved manner also has ramifications for our understanding of how stellar populations vary in their compositions across galactic scales as well as the kinematic evolution and dynamical state of galaxies.
% Where and when stars form is a crucial ingredient that must be firmly constrained in order to improve our grasp on how the physical properties of the ISM not only trigger/curtail star formation but also how they influence the dynamical evolution and kinematics of the galaxy.
% the phenomena of star formation and feedback, and their impact on the fate of the baryons over cosmic time.

% Our understanding of the structure, composition and evolution of galaxies has strongly improved in the last decades, mostly due to new results based on large spectroscopic and imaging surveys. In particular, the nature of ionized gas, its ionization mechanisms, its relation with the stellar properties and chemical composition, the existence of scaling relations that describe the cycle between stars and gas, and the corresponding evolution patterns have been widely explored and described. More recently, the introduction of additional techniques, in particular Integral Field Spectroscopy, and their use in large galaxy surveys, have forced us to re-interpret most of those recent results from a spatially resolved perspective.

It has long been known that properties of the birth sites of stars can differ substantially between regions not only within a galaxy, but also between galaxies of different global phenotypes (masses, morphologies, etc.). Across large parts of a galaxy, star-forming gas can show a range of densities and metallicities, often correlated with the environment and changing with galactocentric radial location \citep[e.g.,][]{pagel81, koda09, heyer15}. Past observations capable of directly probing cloud-scale quantities like velocity dispersions and surface densities for cold gas - such as those with the Atacama Large Millimeter/submillimeter Array (ALMA) and the Northern Extended Millimeter Array (NOEMA) - have indeed revealed sizeable deviations amongst the properties of star forming molecular clouds in different sites, namely, the Galactic centre \citep{shetty12, battersby20}, outer parts of our Galaxy \citep{rice16, md16} and local star-forming galaxies \citep{schinnerer13, sun18}. Along the same lines, merging and starburst galaxies have been shown to boast higher densities, line-widths and diluted metallicities \citep{cortijo17, elmegreen16, irwin94} than most other environments. Nevertheless, results in this area have been limited to a handful of physical parameters, and to small samples of galaxies lacking diversity in their global properties \citep{leroy08, bolatto08, wong11, druard14, faesi18}.

In the last decade, our understanding of both gas and stellar properties on resolved ($\sim$kpc) scales has greatly benefitted from the advent of integral field spectroscopy. Owing to the use of wide-field multiplexed integral field units (IFUs) in large galaxy surveys - namely, the Calar Alto Legacy Integral Field Area Survey (CALIFA; \citealt{sanchez12}), the Sydney-AAO Multi-object Integral field spectrograph Galaxy Survey (SAMI; \citealt{croom12}), and Mapping Nearby Galaxies at Apache Point Observatory survey (MaNGA; \citealt{bundy15}) - we now have simultaneous photometric+spectroscopic information across relatively large radial extents of low-redshift galaxies for a statistically significant sample with varied structural and environmental properties. Recent works from these surveys have explored local ionization states of galactic regions and the different processes responsible for them, the presence of resolved scaling relations down to kpc-scales, their comparison with global counterparts, and the role of star-formation and dynamical processes in giving rise to these relations \citep[see][and references therein]{sanchez20a, sanchez20b}. Furthermore, the combination of IFU surveys with millimeter-wave interferometry, e.g., EDGE-CALIFA\footnote{Extragalactic Database for Galaxy Evolution Survey-CALIFA} \citep{bolatto17} and ALMaQUEST\footnote{ALMA-MaNGA QUEnching and STar formation} \citep{lin19} has now poised us to better understand the kinematical properties of molecular gas and its role in the fueling and quenching of star formation on resolved scales.

On the theoretical front, to investigate star formation and its effects at a higher level of spatial detail, intermediate-scale simulations capable of resolving supernova feedback (for e.g., the TIGRESS framework by \citealt{kimostriker17}; see also \citealt{gatto17, kannan20}) are now helping to close the gap between stellar ($\lesssim$ pc) and cosmological scales ($\gtrsim$ Mpc). In general, these are vertically stratified ``tall-box" simulations representative of specific local star-forming regions within a galaxy with domain sizes of $\sim$kpc and $\sim$pc-scale resolution. In such simulations, the adopted models allow for a more comprehensive and self-consistent evolution of a self-gravitating multiphase ISM with an explicit treatment of star formation and supernova feedback and very few \emph{a-priori} assumptions. The ISM content and disc gravity within such a framework are parameterized by a set of physical properties (e.g., gas/stellar surface density, dark matter density, gas metallicity, stellar scale height/vertical velocity dispersion etc.) which are representative of the patch being simulated, and play a central role in governing the process of star formation. 
%As an example, while surface densities dictate the availability (or lack) of pressure support for the collapsing gas, metals control gas cooling and provide sites for formation of dust/molecules in the ISM. 
% As an example, while gas surface density modulates the amount of fuel available for star formation, the combination of stellar surface density and scale height dictates the gravity experienced by the collapsing gas and in turn influences how fast star formation would occur \citep[e.g.,][]{bolatto17,shi18}.
Hence, performing a systematic exploration of these parameters in these simulations comprises a vital effort towards achieving a thorough insight into how diverse galactic environments influence the formation of stars, and consequently stellar feedback and outflow properties. Unfortunately though, given the high dimensionality of this parameter space, and the immense computational cost involved in conducting a sweep through it, this problem does not lend itself well to a brute-force approach and requires a better sampling scheme to pick out the most essential initial conditions.

As part of the SMAUG consortium\footnote{\url{https://www.simonsfoundation.org/flatiron/center-for-computational-astrophysics/galaxy-formation/smaug}{}}, we undertake in this paper the task of generating and statistically surveying the multi-dimensional parameter space of the aforementioned local physical properties of star-forming sites in a large-volume cosmological simulation. Specifically, we use the IllustrisTNG50 simulation to do a coarse-grained ($\sim$kpc-scale) exploration of the gas surface density, gas metallicity, stellar surface density, stellar vertical velocity dispersion, epicyclic frequency, and dark-matter volumetric density in a statistically significant sample of galaxies across a wide range of galaxy mass as well as redshift. While these properties may not potentially be exhaustive in describing the conditions of star formation - either because other properties have a less obvious connection to star formation or because the timescales for their evolution are short compared to the star formation time - we have aimed to examine the most common quantities upon which contemporary first principle based models are built (Table \ref{table:params} summarises the list). We study these physical parameters in the context of ongoing star formation (via star formation surface density) and the galactic location/environment (radial galacto-centric distance) of each site. In doing this study, our goal is to uncover the region(s) of the parameter space over which most star-forming environments exist, or in other words, identify the birth-conditions in which most stars in the universe form. Alongside this, our work will enable the recognition of meaningful initial conditions for future local/tall-box simulations, and help devise an optimal strategy for their exploration. Doing so would enable us to understand the link between star formation and outflows directly, which is a key element of the SMAUG project's larger goal to develop and implement advanced sub-grid recipes, such as for star-formation, winds and black hole feeding, that are built on the results of simulations that explicitly resolve small-scale physics, thereby eliminating the necessity of tuning to observations \citep[see][for other work in SMAUG]{smith20, kim20, angles-alcazar20, li20a, li20b, fielding20, pandya20}. Lastly, the results of our work will also be a useful research tool for pursuing in detail future lines of inquiry including but not limited to investigating the universality of resolved scaling relations in galaxies, and the intricate link between local star-formation histories and the scaling laws that control star formation within galaxies.

This paper is organised as follows: Section \ref{sec2} provides a brief description of several features of the IllustrisTNG simulations that are most pertinent to this study. Section \ref{sec3} describes our detailed methodology for generating the ISM physical parameter space. In Section \ref{sec4}, we present a picture of the parameter space in one dimension, focusing mainly on the distributions of all properties, as well as their dependence on galaxy masses and cosmic time. Section \ref{sec5} portrays a multi-dimensional view of all of the parameters, where we briefly explore resolved scaling relationships amongst properties, and later, describe the results of dimensionality reduction conducted on the hyper-parameter-space. In Section \ref{sec6}, we make a qualitative comparison of our simulation results with resolved observations from the MaNGA IFU survey and report our findings. Finally, we summarise the conclusions of our work in Section \ref{sec7}.

\section{The TNG Simulations}\label{sec2}
% \label{sec:illustristng} % used for referring to this section from elsewhere
The IllustrisTNG project is a suite of gravo-magnetohydrodynamic simulations \citep{marinacci18, nelson18, springel18, naiman18, pillepich18b, tngrelease} consisting of three separate cosmological volumes (TNG300, TNG100, and TNG50) run using the moving-mesh code \texttt{AREPO} \citep{springel01} at distinct mass resolutions. The physical model employed is described in \citealt{pillepich18} , and includes subgrid treatments of star formation \citep{sh03}, metal enrichment from stellar evolution \citep{naiman18}, ideal magneto-hydrodynamics \citep{pakmor13}, SN-winds and AGN feedback \citep{weinberger17}, and has been shown to yield results that agree with observations over a diverse range of galaxy properties. For this work, we primarily use the highest resolution realisation of the smallest volume box in the suite, i.e. TNG50-1.  Here, we present a brief summary of the key features of TNG50-1 (hereafter identified as TNG50; \citealt{pillepich19, nelson19}), and refer the interested reader to the papers mentioned in this section for further details on the TNG suite.

TNG50 has a uniformly-sampled domain volume of roughly 50$^3$ Mpc$^3$ with 2 $\times$ 2160$^3$ initial resolution elements, and mass resolution of 8.5 $\times$ 10$^4$ and 4.5 $\times$ 10$^5$ solar masses for baryons and dark matter respectively. The co-moving gravitational softening lengths for dark matter and collisionless star particles is 290 pc, while the gas gravitational softening length is adaptive and set by its cell size, with a floor at 74 pc. Both of these values are considerably smaller than the analysis scale we are interested in (i.e. $\sim$kpc), and hence ensure ample resolving power. At these values, TNG50 provides an exceptional combination of volume alongside resolution that allows us to meaningfully investigate spatially-resolved star formation in this study.

\subsection{Star Formation in TNG50}\label{sec2.1}
The process of star formation from dense gas in all TNG simulations is governed by an updated \citet{sh03} \citepalias[hereafter][]{sh03} sub-grid model of star formation, which uses a specified density threshold as a criterion for star formation to set in. Gas less dense than the threshold value $n_{\rm th} = 0.13$ cm$^{-3}$ (in physical units) is considered non star-forming, and its behavior is driven purely by hydrodynamics (in addition to gravity) based on an ideal-gas equation of state. Whereas, above this density value, the model treats the inter-stellar medium as an admixture of two phases of gas i) a cold, dense star-forming cloud phase, and ii) a hot, ionised phase. Of these, the star-forming phase stochastically turns into stars on a timescale set by the local cold gas density. This conversion of gas into stars is then regulated by the heating of the ISM due to supernova feedback (assumed to be instantaneous in this case) resulting from the death of a fraction of the formed stars. Finally, in this self-regulated regime, the model describes the bulk properties of this multiphase high-density gas, such as its pressure and temperature, in terms of an effective equation of state as a function of density.

\begin{deluxetable*}{l c c}
\tablenum{1}
\label{table:params}
\tablecaption{Constituents of the resolved ISM parameter space measured in this work, their physical importance, and their units of measurement. \label{table1}}
\tablewidth{0pt}
\tablehead{
\colhead{Property} & \colhead{Physical Role} (in tall-box models) & \colhead{Units} 
}
\startdata
Gas surface density, $\Sigma_{\mathrm{gas}}$ & Self-gravity & M$_\odot$ ${\rm kpc}^{-2}$\\ 
Stellar surface density, $\Sigma_{\mathrm{\star}}$ & External gravity & M$_\odot$ ${\rm kpc}^{-2}$\\
Dark matter volumetric density, $\rho_{\mathrm{DM}}$ & External gravity & M$_\odot$ ${\rm kpc}^{-3}$\\
Stellar vertical velocity dispersion, $\sigma_\star$ & External gravity & km s$^{-1}$\\
Epicyclic frequency, $\kappa$ & Gravitational shear & km s$^{-1}$ kpc$^{-1}$\\ 
Gas metallicity, $Z_{\mathrm{gas}}$ & Gas cooling + heating & dimensionless\\
Star formation rate surface density, $\Sigma_{\rm sfr}$ & -- & M$_\odot$ yr$^{-1}$ kpc$^{-2}$\\
Galactocentric radius, $R$ & -- & kpc
\enddata
\tablecomments{All spatial quantities here are expressed in physical units (not co-moving), and property roles are defined in the context of their contribution to the tall-box simulation physics.}
\end{deluxetable*}

The key way in which the model incorporated in the TNG simulations differs from the original \citetalias{sh03} model is in its use of a different initial mass function (\citealp{chabrier03} instead of \citealp{salpeter55}) as well as a softer equation of state. Additionally, on account of the numerical challenge of forming stars due to the very high resolution and consequently extremely short MHD timesteps in the TNG50 simulation, the model implemented therein employs a steeper relationship between the equilibrium star formation timescale and gas density ($t_{\star} \propto n^{-1}$) for the densest gas in the simulation ($\gtrsim 230$ $n_{\rm th}$) as opposed to the canonical scaling ($\propto n^{-1/2}$) set by the observed Kennicutt-Schmidt relation \citep{schmidt59, kennicutt98} used in other TNG boxes. This change was made for numerical efficiency reasons alone, and was found to have no significant impact on the overall gas content or galaxy properties in the simulation (for more details, see \citealp{nelson19}).

Utilizing the TNG simulations, a number of studies seeking to understand star formation and its implications have reported promising outcomes. E.g., \citet{tacchella19} investigate the connection between galaxy morphology and star formation, finding that the morphology of galaxies exhibits only a weak correlation with their star formation activity, which matches the observed correlation. Additionally, \citet{donnari19} have characterised the star formation activity of simulated galaxies, and shown that the slope and normalisation of the star-forming main sequence in TNG are in excellent agreement with observations at $z = 0$. Using an improved framework to estimate neutral gas abundances, \citet{diemer19} demonstrated a good agreement between the simulated and observed galaxy HI size-mass relationship as well as the overall gas fractions. And most recently, \citet{nelson19} have described how the subgrid input parameters in TNG successfully give rise to a realistic multi-phase structure and diverse properties of feedback-driven galactic outflows. These studies, combined with the multitude of results on other aspects of galaxy formation, provide a solid empirical validation to the TNG model.

\section{Methods and Analysis}\label{sec3}
\subsection{Galaxy Sample Selection}\label{sec3.1}
The large volume of TNG50 provides a statistically sizeable sample of galaxies at a resolution capable of discerning the internal structure of galaxies. Haloes and their substructure are identified in the simulation using the standard Friends of Friends (\texttt{FoF}; \citealp{turner76}) and \texttt{SUBFIND} \citep{springel01, dolag09} algorithms. The FoF algorithm identifies collective groups of dark matter particles (aka halos) based on their physical proximity, whereafter the SUBFIND algorithm identifies gravitationally self-bound associations of all resolution elements combined within each halo (aka subhalos). The gas, stars, black holes and dark matter associated with the most massive subhalo within a FoF halo are considered as belonging to the \textit{central} galaxy, while the rest of the self-gravitating substructures, when present, are classified as \textit{satellite} galaxies. At the present epoch, the simulation contains a total of $\sim$96,000 galaxies with M$_\star \gtrsim 10^5$ M$_\odot$ with a variety of morphologies, sizes and formation histories \citep[e.g.,][]{pillepich19, joshi20, pulsoni20}. For our specific analysis, we construct a sample consisting of both centrals and satellites as follows:
\begin{enumerate}[leftmargin=*]
    \item We select galaxies with stellar masses M$_{\star}$ in the range $10^{7-11} {\rm M}_{\odot}$ at redshifts $z$ = \{0, 0.5, 1, 2, 3\}. The lower limit of $10^7$ M$_\odot$ is chosen to ensure measurement robustness in view of the mass resolution of the simulation. In this mass range, we find a range of morphologies, from the extended, actively star-forming discs to quenched (non star-forming) elliptical systems, similar to those observed in the real universe.
    
    \item For the selected galaxies, we implement a minimum cut on the number of particles contained within the galaxy to be at 100 each for the gas, stars and dark matter, such that all three components of the galaxies are sufficiently resolved. At the standard TNG50 resolution, this corresponds to a minimum M$_\star \simeq 10^{6.9}$ M$_\odot$ and $M_{\rm dm} \simeq 10^{7.7}$ M$_\odot$. We also impose a minimum total star formation rate threshold of 5 $\times 10^{-4}$ M$_\odot$ yr$^{-1}$ within a spherical volume of twice the 3D stellar half-mass radius ($R^\star_{1/2}$) centered at the centre of mass of each galaxy. Even though the latter measure leads to an exclusion of galaxies that are fully or almost-fully quenched, it does not adversely impact this study given that such galaxies are not expected to contribute much in terms of actively star forming regions.
    
    \item Lastly, due to our inclusion of satellite galaxies in the sample, we take precaution to remove any misidentified subhalos such as clumps or fragments in the outer parts of a halo that did not arise from standard cosmological processes of structure-formation and collapse, using the \texttt{SubhaloFlag} in the simulation (see \citealt{tngrelease} for more details).
\end{enumerate}

Finally, our sample contains 10394, 13806, 16663, 21039 and 20630 galaxies respectively at $z$ = \{0, 0.5, 1, 2, 3\}. In totality, our selected sample at each of the redshifts encompasses $>$80\% of the total instantaneous star formation occurring in the simulation volume. For most of the analysis in this paper, we will focus on $z = 0$ galaxies with stellar mass in the range $10^{9-10} {\rm M}_{\odot}$, but also explore the variation in resulting trends with galaxy stellar mass and redshift.

\subsection{Galaxy Data Processing}\label{sec3.2}

Since our goal is to characterise the birth environments of stars in the context of their corresponding ISM properties, we set up a multi-dimensional physical parameter space by conducting a coarse-grained measurement of these properties from individual star-forming regions within our galaxy sample. In a nutshell, we divide each galaxy into spatially-resolved regions by projecting it onto a 2-dimensional image grid (in the manner of \citealp{diemer18}), where each pixel represents an ISM patch, and subsequently extract the physical parameters of interest as column-integrated quantities from the image pixels so produced. We further elaborate on the principal facets of our analysis below.\\

\underline{Particle Smoothing:} Due to the discretised nature of the simulation volume and a finite mass resolution, each cell/particle in it represents an unresolved entity that should ideally be spatially distributed. For investigating spatially resolved properties, as we do in this paper, it is thus important that we alleviate the effects of this coarse sampling so as to avoid biasing our quantitative analysis or making it dependent on the choice of the analysis scale adopted. 

In this work, we utilise a smoothing scheme where we re-sample each star, gas and dark matter particle such that it gives rise to a finitely extended distribution of \emph{sub-particles}. The smoothing length $\sigma$, which governs the spatial extent within which the sub-particles originating from a parent particle are distributed, is determined adaptively for each particle based on the local density of the corresponding particle type. This translates into the radius encompassing a fixed number of nearest neighbours $N_{\rm ngb}$, which we choose to be 32 and 64 for stars and dark matter respectively. For gas, we use one-third of the distance to the 32nd nearest neighbour. We note that our choice of $N_{\rm ngb}$ here is determined based on a visual conciliation between noticeable pixelation effect and fading of resolved structure in galaxies. Using this smoothing length as the standard deviation, we then convolve each particle with a discrete 2D Gaussian kernel of size 6$\sigma$ on a side and resolution 1 kpc centred on its position, giving us a collection of sub-particles that inherit the physical properties of their parent particle in a manner that conserves the extensive properties (mass, energy, momentum) of the parent particle. From here on, we use these re-sampled sub-particles in lieu of the original particles for further analysis.\\

\begin{figure*}[t!]
\centering
\includegraphics[width=0.9\textwidth]{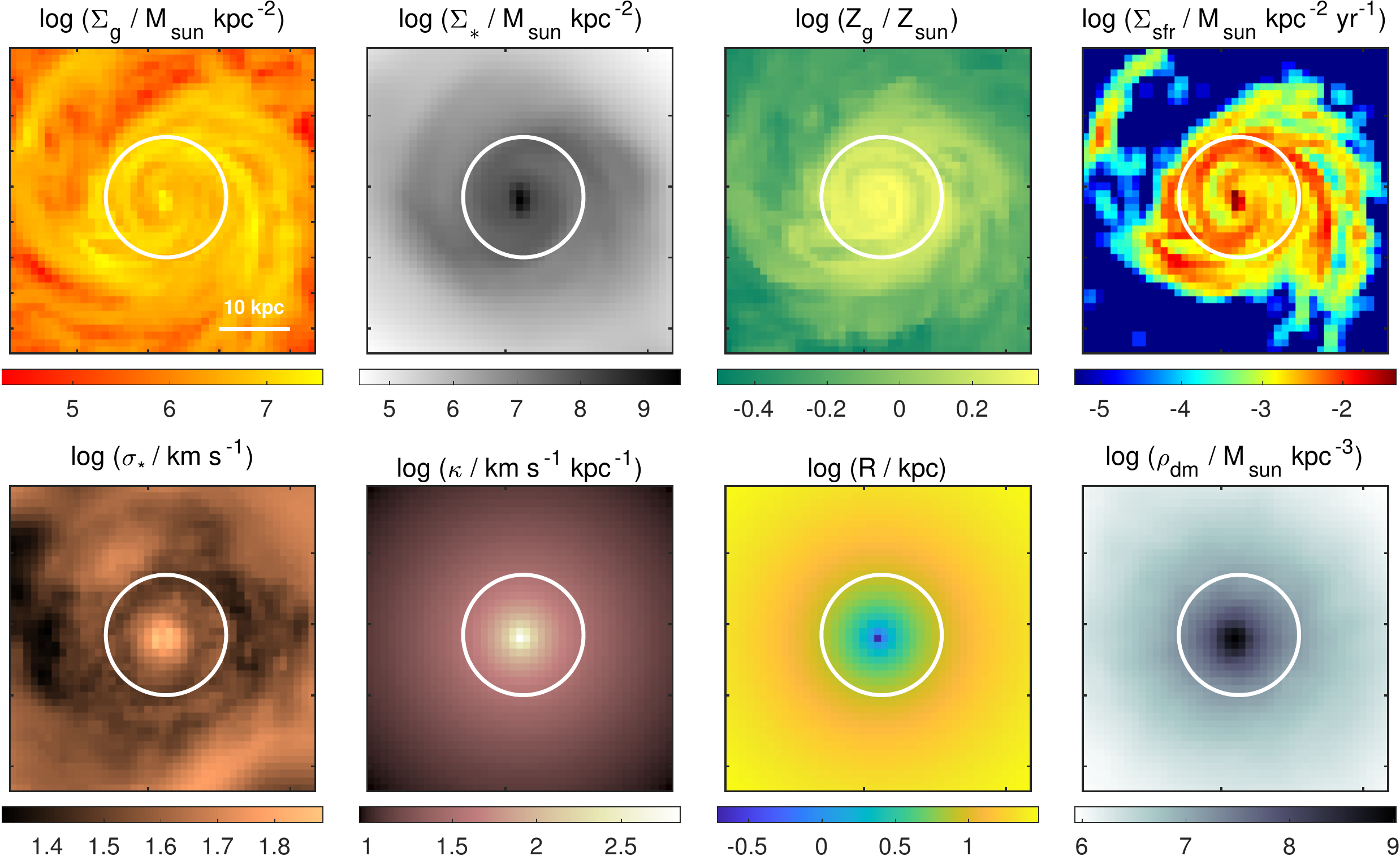}
\caption{Projections of physical properties for an example disc galaxy at $z = 0$ with M$_{\ast}$ = 2.84 $\times 10^{10} \textrm{ M}_\odot$ in TNG50. The images are $\approx$43 kpc on a side, which is twice the radius encompassing 95\% of all star-forming gas particles in this galaxy. The white circles denote twice the stellar half-mass radius at $\approx$8.5 kpc.}
\label{fig:fig1}
\end{figure*}

\underline{Galaxy rotation and coarse-graining:} For each galaxy, we first compile the list of both the parent (original) gas/star/dark-matter and corresponding sub-particles comprised within it. We then calculate the angular momentum of the galaxy in its centre of mass rest frame, taken here to be that of all the star-forming \emph{parent} gas particles within 2$\times R_{1/2}^{\star}$, and use it to perform a rotation on the galaxy such that the cartesian z-axis is aligned with the direction of the calculated angular momentum vector. This operation transforms the galaxies with(without) rotational symmetry to a face-on(random) orientation, which is then spatially binned using a grid of predetermined size and resolution along the x-y plane to create an image representation for each of the desired physical quantities. In this study, we choose a fixed pixel scale (grid resolution) of 1 kpc, which closely emulates the domain size of local ISM simulations, as well as the sampling scale of modern IFU surveys. The overall size of the square image is given by $L_{\rm image} \approx 2 \times \mathrm{max}(2R_{1/2}^{\star}, R_{\rm 95, sfr})$, where $R_{\rm 95, sfr}$ is the radius within which 95\% of all star-forming parent gas particles of the galaxy are enclosed. This criterion allows us to include most of the star formation in each galaxy while also accounting for a significant fraction of its visible stellar component. A discussion on convergence with different values of pixel scale and simulation resolution is presented in Appendix \ref{appendixA}. Since we do not impose any restrictions on $\mathrm{max}(2R_{1/2}^{\star}, R_{\rm 95, sfr})$ to assume an integer or half-integer value, we do not expect the center of the image grid to coincide with the coordinate center of the galaxy.

\subsection{Generation of Property Maps}\label{sec3.3}
To obtain the desired physical property maps for a galaxy, we utilise the correspondingly binned gas, star and dark matter sub-particles gravitationally bound to the subhalo (as determined by \texttt{SUBFIND}) that lie within a vertical column of height z = $\pm$20 kpc (unless otherwise noted) relative to the projection plane. This value is arbitrarily chosen, and was selected to minimise the contamination from the hot gaseous corona as well as from halo stars, while preserving the contribution from the diffuse stellar component associated with disc galaxies as well as accommodating galaxies that do not have a well-defined rotation axis. Below, we provide the exact definitions used to calculate local ISM properties of individual pixels (also see Table \ref{table:params} for a summary list):\\

\textbf{Gas surface density $\Sigma_{\rm g}$.} Sum of the masses of all gas (sub-)particles contained within the column divided by the area of the pixel. For non star-forming gas, we only include the fraction of mass present as neutral hydrogen (although this generally includes HI + H$_2$, the simulation does not differentiate between the two). For gas cells below the density threshold, the simulation computes this fraction using the atomic network of \citep{katz96} and a density-based self shielding prescription \citep{rahmati13} in the presence of a time-dependent uniform ultraviolet background \citep{faucher09}.\\

\textbf{Stellar surface density $\Sigma_{\star}$.} Sum of the masses of all star particles present within the column divided by the pixel area.\\ 

\textbf{Gas metallicity $Z_{\rm g}$.} Mass-weighted mean metallicity of the same gas as used for the calculation of $\Sigma_{\rm g}$.\\

\textbf{Star formation surface density $\Sigma_{\rm sfr}$.} Sum of the star formation rate of all gas particles contained within the column divided by the area of the pixel.\\

\begin{figure*}[t!]
\centering
\includegraphics[width=\textwidth]{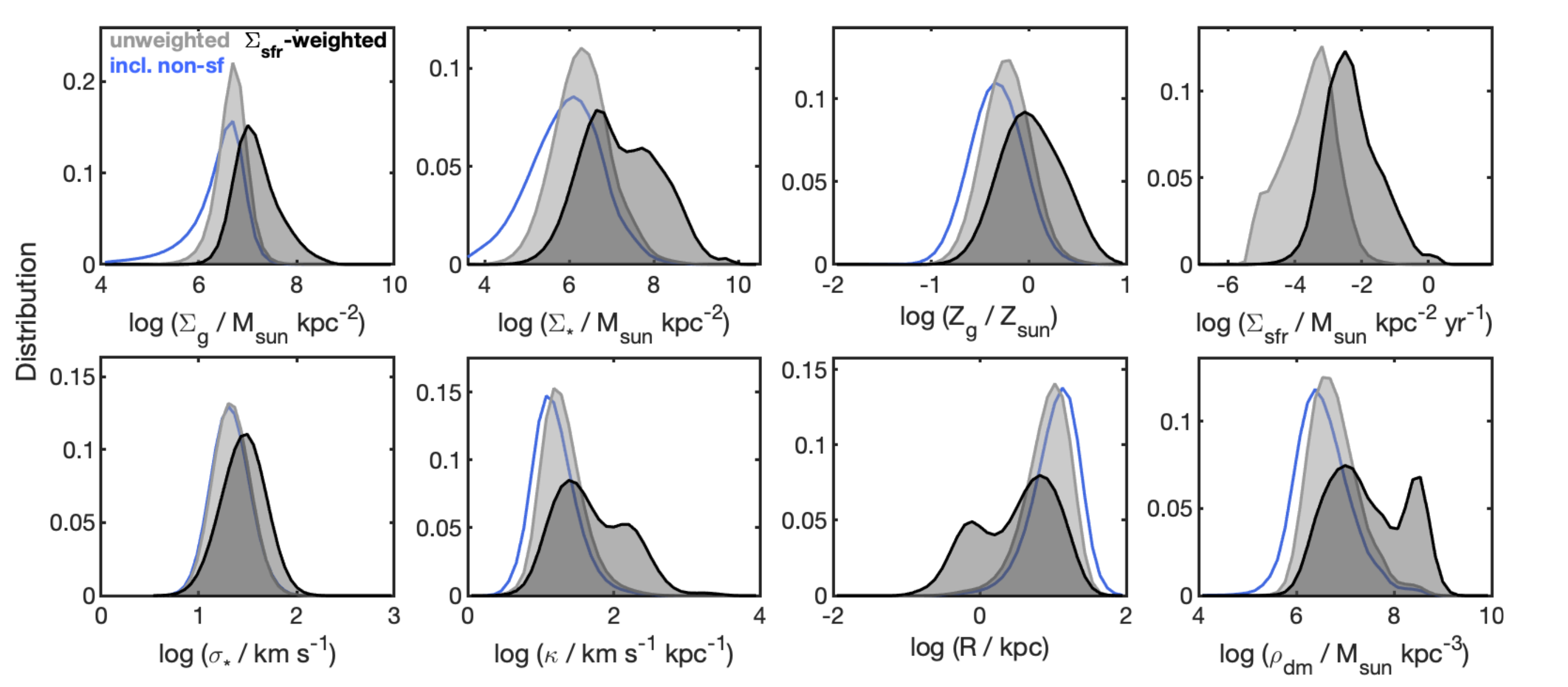}
\caption{Independently normalised 1D probability distributions for the measured physical properties of all star-forming pixels from the sample of $\approx$1900 galaxies with M$_{\star} = 10^{9-10} \textrm{ M}_\odot$ at $z = 0$ in TNG50. Grey curves depict the unweighted distributions for each property using only the pixels/regions with finite amount of star formation in them while black curves depict the corresponding $\Sigma_{\rm sfr}$-weighted distributions. For completeness, we also show in blue the unweighted property distributions obtained from the full pixel dataset inclusive of non-star-forming pixels.}
\label{fig:fig2}
\end{figure*}

\textbf{Stellar vertical velocity dispersion $\sigma_{\star}$.} Mass-weighted standard deviation of the z-velocity of all the star particles present within the column.\\

\textbf{Galacto-centric radius $R$.} Measured based on the number of star-forming gas sub-particles ($N_{\rm sfg}$) inside the pixel. For pixels with $N_{\rm sfg} \le 1$ , $R$ is assigned to be the Euclidean distance between the galaxy centre and the centre of the pixel, whereas when $N_{\rm sfg} > 1$, $R$ is the mean euclidean distance between the galaxy centre and the SFR-weighted mean 2D position coordinates of the gas sub-particles in the pixel.\\

\textbf{Epicyclic frequency $\kappa$.} Calculated using the following expression (simplified from Eq. 3.83-84 in \citealt{binneytremaine08}) involving the galactocentric radius of the pixel $R$ and the circular velocity at that location $v_{\rm c}(R)$:
\begin{equation}
    \kappa(R) = \sqrt{2\left(\frac{v_{\rm c}(R)}{R}\right)^2 \left(1 + \frac{R}{v_{\rm c}(R)}\frac{\partial v_{\rm c}}{\partial R}\right)}
\end{equation}
Here, $v_{\rm c}(R)$ is due to the total mass enclosed within a spherical volume of radius $R$, and defined as $\sqrt{GM(R)/R}$.\\

\textbf{Dark-matter volumetric density $\rho_{\rm dm}$.} Sum of the masses of all dark-matter sub-particles within a column of height z $= \pm h_{\rm \star, z}$ divided by the volume of the column. Here, $h_{\rm \star, z}$ denotes the stellar half-mass height associated with the pixel.\\

As an example, we show in Figure \ref{fig:fig1} images of the aforementioned physical properties for a galaxy with high gas-mass fraction. After generating such images for our entire sample of galaxies, we record the values for all pixels obeying $\Sigma_{\rm sfr} > 0$ (hereafter dubbed `star-forming regions') to obtain the final 8D parameter-space. We note that due to the presence of out-of-equilibrium and merging galaxies in the simulation volume, not all star-forming pixels in our sample are inherited from dynamically stable discs. Nevertheless, we do not exclude such pixels from our analysis as we are interested in exploring all types of star-forming environments in this study. We also, once again, remind the reader that there are potentially additional local properties that may be important in describing star formation, but in choosing the aforementioned quantities, we have attempted to identify those that are commonly discussed in the context of star formation on kpc-scales. It is indeed expected that not all of these quantities would be mutually independent and encode unique information, and we address this aspect in a later section of this paper (see \S\ref{sec5.1}).

\section{Physical Properties of Star-forming Regions}\label{sec4}
One of the primary goals of our study is to understand the dominant regime of star formation in the universe by way of characterising the underlying physical properties of the ISM. A starting point for doing so would be to summarise the statistical characteristics of the properties themselves measured from our large sample of galaxies. Thereafter, one can glean insight into the physical processes driving the shapes of the distribution functions, as well as parse the degeneracies between them, by exploring how the distributions evolve as galaxies evolve in the redshift and stellar mass space.

\begin{figure*}[t!]
\centering
\includegraphics[width=\textwidth]{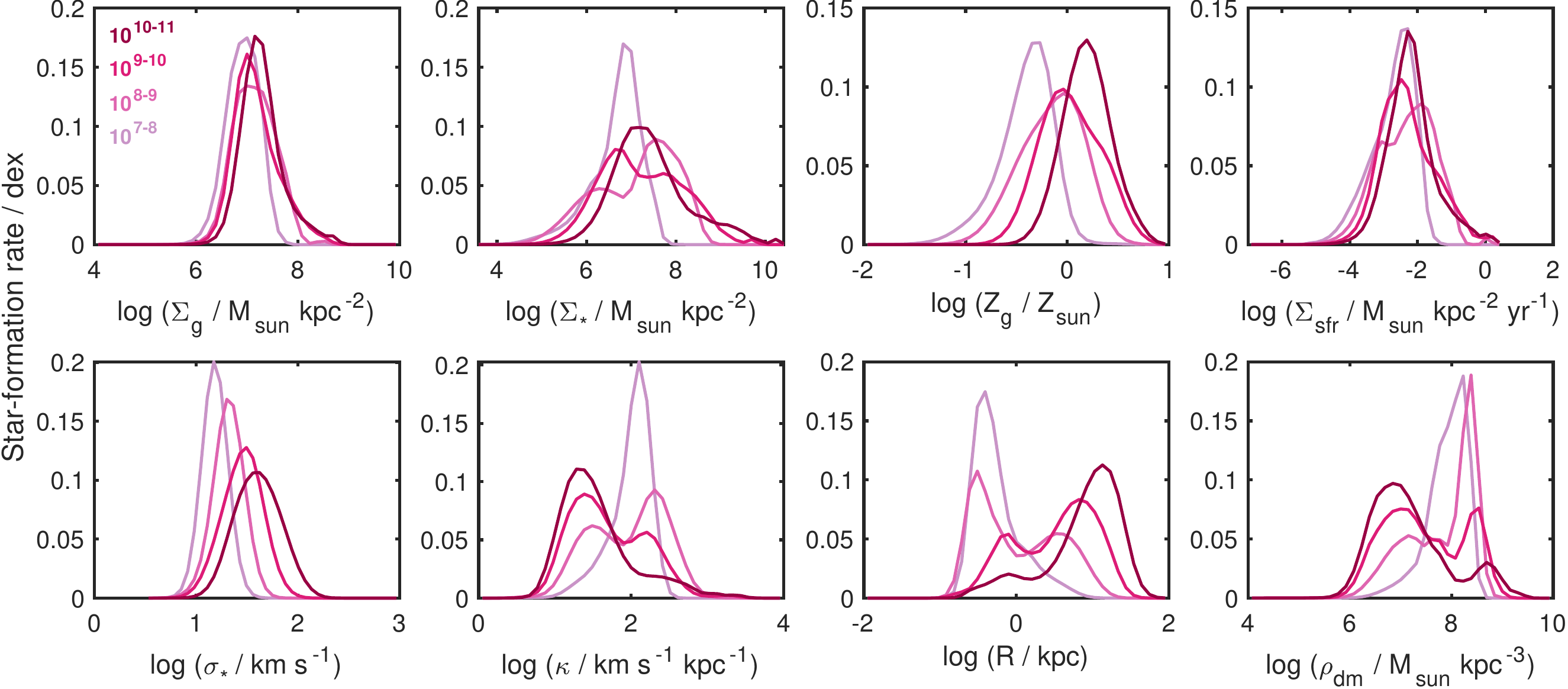}
\caption{Progression in the $\Sigma_{\rm sfr}$-weighted probability distributions of all physical quantities with the stellar mass of the parent galaxy at $z = 0$ in TNG50. Since the curves in each panel are independently normalised, the respective areas under the curve here do not convey the relative amount of star formation occurring in different mass galaxies.}
\label{fig:fig3}
\end{figure*}

Thus, in this section, we begin by presenting probability distributions for the measured physical properties weighted by star formation and highlight their salient features in \S\ref{sec4.1}. We then explore how these distributions change as a function of stellar mass and redshift of the parent system in \S\ref{sec4.2} and \S\ref{sec4.3}. We also present composite distributions of ISM conditions that have given rise to stars throughout cosmic time in \S\ref{sec4.4}. Lastly, in \S\ref{sec4.5}, we investigate the underlying origin of the bimodality seen in the distribution functions presented in \S\ref{sec4.1}.

\subsection{Distribution Functions for Low-redshift Galaxies}\label{sec4.1}
% As noted in our introduction, by virtue of its complex hydrodynamical, radiative, gravitational and magnetic interactions, the ISM can substantially differ in its physical and chemical properties from region to region, galaxy to galaxy. Since these processes also control the rate of star formation, the diversity in these properties in turn reflects on the way star formation is distributed not just spatially, but also across the spectrum of different physical environments in the universe. 

In order to discern which regions of the ISM physical parameter space support most of the overall amount of star formation, it is instructive to look at the distributions of these parameters weighted by the star formation rate of the corresponding pixels. Due to the fixed physical size of all pixels, this is equivalent to weighting by $\Sigma_{\rm sfr}$. In Figure \ref{fig:fig2}, we show independently-normalised one-dimensional distributions of the properties of all star-forming regions belonging to our fiducial sample (which are galaxies with M$_\star = 10^{9-10} {\rm M}_\odot$ at $z$ = 0). The grey curve in each panel depicts the \emph{unweighted} distribution, and in black we show the same distributions weighted by $\Sigma_{\rm sfr}$. We observe that the weighted radius distribution prefers lower values while all other distributions are shifted towards higher values compared to their corresponding unweighted counterparts. This trend is reflective of the intuitive notion that the denser, inner regions of galaxies are more conducive to the formation of stars on account of the gas being dense enough to cool and collapse. More notably, we find that unlike the unweighted distributions, many of the weighted distributions -- all except gas surface density and stellar vertical velocity dispersion -- exhibit a strong bimodality. This suggests that star formation in our sample of galaxies is neither agnostic to the properties of the ISM nor does it favour a specific range of values, but instead preferentially occurs in two distinct environmental regimes. Later in this section, we investigate the origin of this feature from radial star formation surface density profiles of the overall population (see \S\ref{sec4.5}).

\subsection{Dependence on Galaxy Stellar Mass}\label{sec4.2}
In view of the large dynamic range of galaxy masses present in the simulation, we now look at how properties of the ISM in star-forming regions differ between galaxies with different stellar masses. Figure \ref{fig:fig3} shows the $\Sigma_{\rm sfr}$-weighted distributions of ISM properties of regions drawn from present day ($z$ = 0) galaxies in four equal-sized bins of galaxy stellar mass. In each panel, the curves become darker with increasing stellar mass from $10^7$ to $10^{11}$ ${\rm M}_{\odot}$. We find three broad features to be apparent.

First, we notice that the distributions of gas surface density, stellar surface density, and SFR surface density all exhibit a subtle shift towards higher densities and SFR values for higher mass galaxies. While this trend is mostly manifest in the tails (particularly, on the high-$\Sigma_{\rm g/ \star/sfr}$ end), the peak values of the distributions do not significantly vary amongst galaxies of different masses. Given that we are solely looking at actively star-forming regions of the galaxies, the concurrence in the behaviours of $\Sigma_{\rm g}$ and $\Sigma_{\rm sfr}$ is consistent with, and ensues from the fact that the rate of star formation in galaxies is fundamentally governed by the density of gas on sub-galactic scales. In line with this, the two are directly related by construction in the star formation model of IllustrisTNG (\S\ref{sec2.1}). Additionally, the lack of strong variation in the resolved star formation rate (alongside gas and stellar density) distributions with galaxy mass confirms that star formation, being an inherently small-scale process, is rather impervious to the overall gravitational potential of the galaxy, but is instead strongly influenced by the local gravity set by $\Sigma_{\rm g}$ and $\Sigma_{\star}$

\begin{figure*}[t!]
\centering
\includegraphics[width=\textwidth]{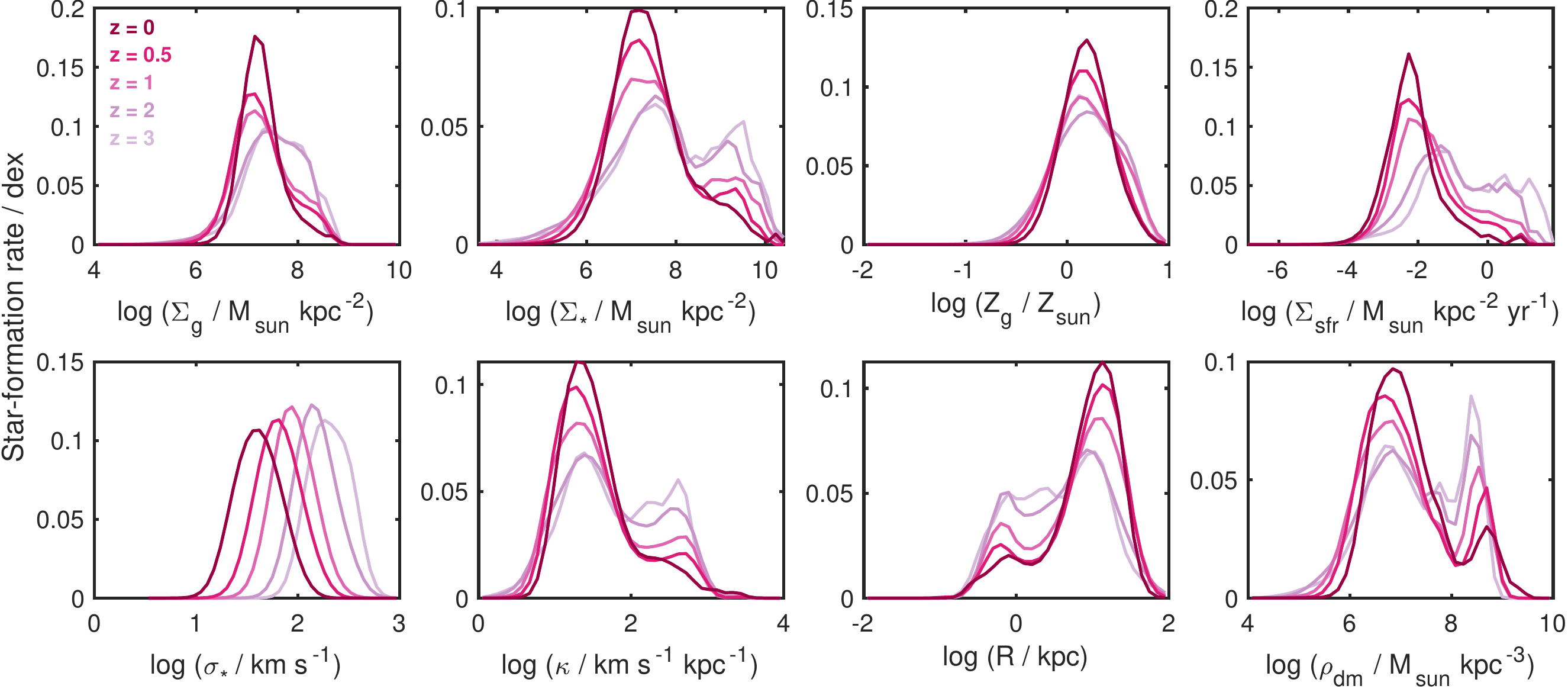}
\caption{Evolution of $\Sigma_{\rm sfr}$-weighted probability distributions of all physical quantities with redshift for the M$_\star = 10^{10-11}{\rm M}_\odot$ galaxy sample in TNG50. All curves shown here are independently normalised.}
\label{fig:fig4}
\end{figure*}

Second, the mean stellar vertical velocity dispersion of star-forming regions monotonically increases as a function of galaxy mass, with the distributions themselves progressively broadening. This dependence can be ascribed to the fact that galaxies with a more massive stellar component require larger dispersions to maintain vertical dynamical equilibrium against the deepening of their gravitational potential wells. Our finding in this case is also corroborated by the results by \citealt{pillepich19}, where they show that the median 3D velocity dispersion and scale height of stellar discs of star-forming TNG50 galaxies indeed increase as a function of mass regardless of the sample redshift. Similarly, the peak gas metallicity also shows an increasing trend with stellar mass \citep{tremonti04}, albeit with a gradual translation of the distribution as a whole to higher values. Star-forming gas in higher mass galaxies is expected to be on average more enriched than in lower-mass galaxies owing to a longer corresponding history of star formation and deeper potential, and consequentially, greater metal production and retention. In contrast, lower mass galaxies have a higher gas fraction relative to their stellar material, thus making the metal content more dilute compared to their high-mass counterparts. Interestingly, the variation in the shapes of $Z_{\rm g}$ distributions closely follows as those of the corresponding $\Sigma_{\star}$ distributions, indicating that the increase in metallicity is systematically linked to the bias towards higher stellar densities in more massive galaxies, hence explaining the presence of a \emph{local} mass-metallicity relationship (cf. \S\ref{sec5.1}).

Finally, from the quantities exhibiting bimodally-shaped distributions, we find that at a given redshift (here, $z = 0$), star formation almost exclusively takes place in the high-DM density innermost regions of low stellar mass galaxies, while in the higher mass galaxies, a gradual suppression of this concentrated star formation paves way for relatively more diffuse star formation in lower density regions. These two regimes are roughly equally populated for galaxies with M$_{\star} \simeq 10^9$ ${\rm M}_{\odot}$, above and below which star formation is prevalent in separate sets of parameters. This trend appears due in part to galaxies being more extended at larger masses, hence availing more area for star formation to happen at large radii. This size increase effect is reflected in the translation occurring in the peak positions of the R distribution towards larger values for larger masses. However, other factors could also potentially be at play, namely, an increasing prevalence of central AGN-feedback in more massive galaxies \citep{bongiorno16, kauffmann03, wang08} as well as mass-dependent secular transformation processes leading to a decline in the central gas supply that give rise to a quiescent dense centre surrounded by a more extended gas-rich annulus where star formation mainly occurs \citep{forbes14a, kormendy12}.

\subsection{Evolution with Redshift}\label{sec4.3}
Having looked at how the local property distributions transform with host galaxy stellar mass, we now explore how these properties vary in similar-mass galaxies as a function of cosmic time. Figure \ref{fig:fig4} shows independently-normalised property distributions for star-forming patches from galaxies at five different epochs from $z = 3$ to present with the host mass fixed in the range $10^{10-11}$ M$_\odot$ at each epoch. The curves get darker towards lower values of redshift. The panels show that the bimodally distributed quantities favour lower density regions at later times compared to regions within similarly massive hosts at higher redshifts, albeit maintaining a similar overall range of values. At fixed mass, galaxies at higher redshifts are more compact potentially giving rise to the aforementioned trend.

\begin{figure*}[t!]
\centering
\includegraphics[width=\textwidth]{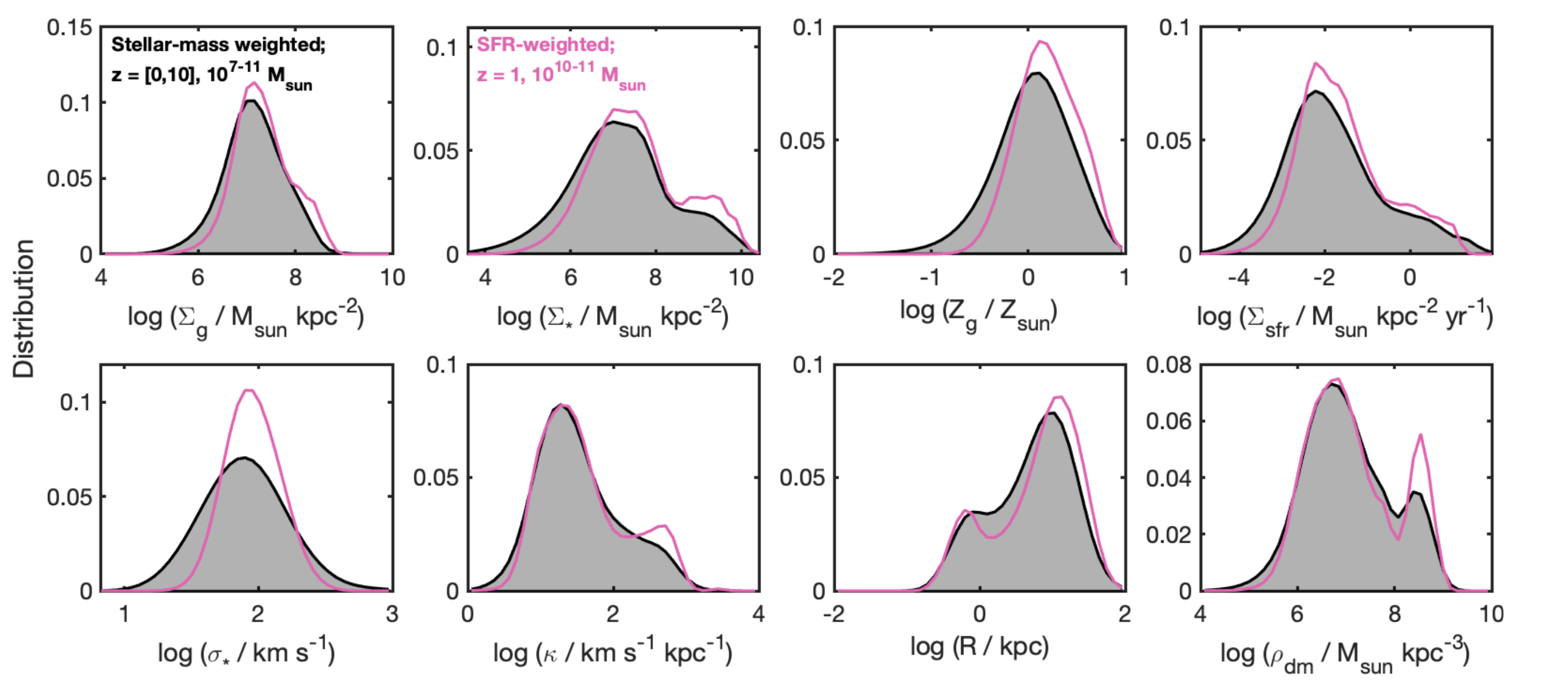}
\caption{Stellar-mass-weighted distributions for the ISM birth conditions in TNG50 for all stars formed during $10 \ge z \ge 0$ in galaxies with stellar mass in the range $10^{7-11}$ M$_\odot$ (black), compared with the corresponding $\Sigma_{\rm sfr}$-weighted distributions of star-forming regions properties in galaxies with M$_\star$ = $10^{10-11}$ M$_\odot$ at $z = 1$ (pink; similar to that in Figure \ref{fig:fig3}). All curves are independently-normalised.}
\label{fig:fig6}
\end{figure*}

Another notable feature appears in the case of stellar velocity dispersion, where a discernible overall shift occurs in the distributions from higher to lower values, while their width remains mostly unaffected. As expected, this trend can be ascribed to the fact that galaxies at low redshifts, especially star-forming, have a higher degree of rotational support and relatively thinner discs.

Lastly, we see that the shape of the metallicity distribution mildly changes from being bimodal for galaxies at high redshifts to unimodal at low redshifts. As evident from the corresponding  $\Sigma_{\star}$ distributions, this pattern can be ascribed to the removal of high-density, high-metallicity gas from the central regions of galaxies. However, the range of local metallicity values for our galaxy sample does not substantially evolve with redshift. Considering that the total gas fraction of galaxies appreciably varies with redshift at fixed stellar mass \citep{santini14}, the constancy in local metallicity distributions demonstrates that the chemical evolution in galaxies is predominantly driven through outflows and less so via gas inflow and stellar evolution \citep{torrey19}. The redshift dependence of the integrated mass-metallicity relationship (MZR) must thus result from galaxy populations losing gas while sampling from an underlying unevolving distribution of local gas metallicity.

\subsection{Star formation across Cosmic Time}\label{sec4.4}
In the preceding subsections, we looked at the full expanse of conditions under which star formation occurs in the universe at fixed time, and explored how these conditions depend on galaxy mass and epoch. It is then natural to ask: what are the distributions of stellar birth conditions that have collectively given rise to all the stars that have ever been formed? To answer this, we now look at resolved ISM property distributions of star-forming regions across a very wide window of cosmic time. In Figure \ref{fig:fig6}, black curves indicate independently-normalised \emph{stellar-mass-weighted} distributions obtained from a composite dataset of pixels from galaxies with M$_\star$ = 10$^{7-11}$ M$_\odot$ at 18 different snapshots - with a roughly uniform spacing of 0.1 in scale factor - between $z$ = 0 and 10. Each set of property values (corresponding to a pixel) is weighted by the mass of newly-formed stars contributed by the associated star-forming region. Assuming that the distribution of star-forming region properties varies weakly enough with time, we calculate the stellar-mass contribution for each pixel to be its instantaneous star formation rate (same as $\Sigma_{\rm sfr}$ due to unit pixel size) times the inter-snapshot duration $\Delta t_{\rm sf}$. More precisely, for snapshot $i$, $\Delta t_{\rm sf}^i =\left(t^{i-1} - t^{i+1}\right)/2$, where $t^i$ is the lookback time associated with snapshot $i$. For the last snapshot (corresponding to $z = 0$), it is $\left(t^{i-1} - t^{i}\right)/2$.

The property distributions have an overall strong resemblance to the distributions we saw in \S\ref{sec4.3} for galaxies with M$_\star$ = 10$^{10-11}$ M$_\odot$ at $z = 1$ (also shown in Figure \ref{fig:fig6} in pink). In all of the panels barring $\sigma_\star$, the similarity is apparent both in the locations of the peaks as well as their amplitudes signifying a connection between local ISM conditions contributing new stellar mass in the universe to the conditions sustaining star-formation in massive galaxies at $z = 1$. Previously published work on the global star formation histories of galaxies and the evolution of star formation efficiencies have shown that galaxies with halo masses in the range $10^{11.5} - 10^{12.2}$ M$_\odot$ have the highest star formation efficiency at every epoch, and are responsible for making most of the stars in the universe \citep{behroozi13b}. From the stellar mass-halo mass relationship \citep{moster10}, this is roughly equivalent to the stellar mass range $10^{10} - 10^{11}$ M$_\odot$, in keeping with our result. Moreover, it has been estimated from both observations and simulations that galaxies in this mass range build up $\sim$ 80-90 \% of their stellar mass at $z \lesssim 2$ (mostly in their discs; \citealp{tacchella19, behroozi13a}) and have a mass-weighted mean stellar age of $\sim$7 Gyrs, corresponding to z $\simeq$ 1 \citep{behroozi13a}. This again, is well-reflected in our current findings. Nonetheless, due to the influence of stellar mass formed along the entire cosmic star formation history - which includes stars formed at earlier times and in galaxies with lower stellar masses - the overall distributions here are somewhat broader than the ones we saw in the preceding sections for fixed mass and time.

\subsection{The Origin of Bimodality}\label{sec4.5}
To obtain some insight into the source of bimodality in several of the ISM properties we have seen thus far, we now examine the spatial distribution of star formation. Specifically, since the bimodality in the distributions arises from weighting by $\Sigma_{\rm sfr}$, we look into how the mutual variation between star formation and ISM properties can generate bimodality in the respective distributions of those properties. In the preceding sections, we saw a concomitant evolution of bimodalities in multiple quantities, hinting at a common underlying reason for their appearance. Based on this, we choose only one of the parameters - the galactocentric radius $R$ - in this section for illustrative purposes, and expect our inferences to apply equivalently to the other bimodally-distributed quantities, namely, $\Sigma_{\star}$, Z$_{\rm g}$, $\Sigma_{\rm sfr}$, $\kappa$, and $\rho_{\rm dm}$.

\begin{figure}[t!]
\centering
\includegraphics[width=0.49\textwidth]{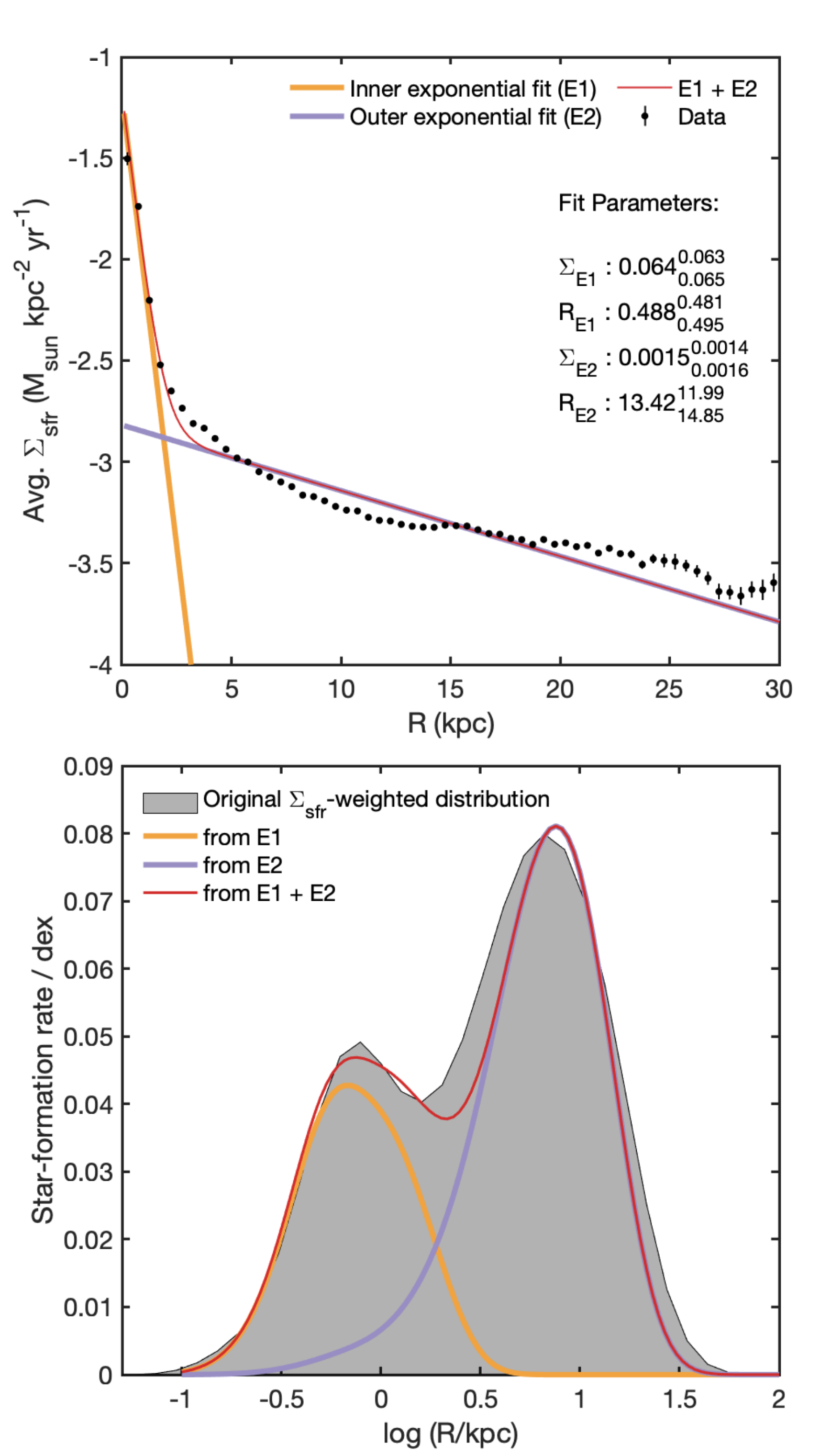}
\caption{Top: A two-component fit to the average radial star formation rate surface density profile (black markers) generated using the sample of pixels with $\Sigma_{\rm sfr} > 0$ belonging to TNG50 galaxies in the mass range 10$^{9-10}$ M$_\odot$ at $z = 0$. The best fit curves are shown in yellow (inner exponential), purple (outer exponential) and crimson (total). Error-bars on the markers indicate relative standard error of mean, and the abscissa limits do not contain data points that were excluded from the fit. Bottom: The two components making up the average star formation rate profile result in a bimodal shape of the galactocentric radius distribution of star-forming regions (cf. panel 7 in Figure \ref{fig:fig2}).}
\label{fig:fig5}
\end{figure}

In Figure \ref{fig:fig5} we show the sample-averaged $\Sigma_{\rm sfr} (R)$ profile constructed using the measured values of $R$ and $\Sigma_{\rm sfr}$ of all star-forming pixels from our fiducial sample of galaxies\footnote{It is important to clarify here that due to the exclusion of pixels with no finite amount of star formation, the profile shown in Figure \ref{fig:fig5} does not represent a true radially-averaged density profile, in a way that would be constructed using a stacked galaxy sample. Here, unlike the total pixel population in a galaxy, the number of star-forming-only pixels per radial bin does not scale with the bin radius in a linear fashion.}. The profile appears to be comprised of two disparate components, which we demonstrate below to be responsible for the two distinct peaks in the distributions of quantities. To ascertain the exact shape of the profile, we fit it with a two-component analytical model consisting of an inner plus an outer exponential component as:

\begin{equation}
    \langle\Sigma_{\rm sfr}(R)\rangle = \Sigma_{\rm E1}e^{-R/R_{\rm E1}} +  \Sigma_{\rm E2}e^{-R/R_{\rm E2}}
\end{equation}

where $\Sigma_{\rm E1}$($\Sigma_{\rm E2}$) are the normalisations, and $R_{\rm E1}(R_{\rm E2})$ are the scale lengths associated with the inner(outer) exponentials. To calculate the fit parameters, we utilise \texttt{MATLAB}'s `trust-region' algorithm (a kind of non-linear least squares formulation) with bisquare weights. The data points used for fitting are obtained by binning the sample values at a uniform interval of 0.5 kpc, while our choice of the model itself is motivated by commonly used profiles in the literature to describe azimuthally-averaged radial distributions of HI gas and star formation rates in galaxies. We exclude from our fitting procedure any data points with relative standard error of mean values exceeding 10\%. Figure \ref{fig:fig5} top panel depicts the fitting procedure results for the fiducial galaxy sample. The overall best fit profile is represented as a solid crimson line with the individual components plotted as yellow (inner exponential) and purple (outer exponential) curves.

\begin{figure}[t!]
\centering
\includegraphics[width=0.47\textwidth]{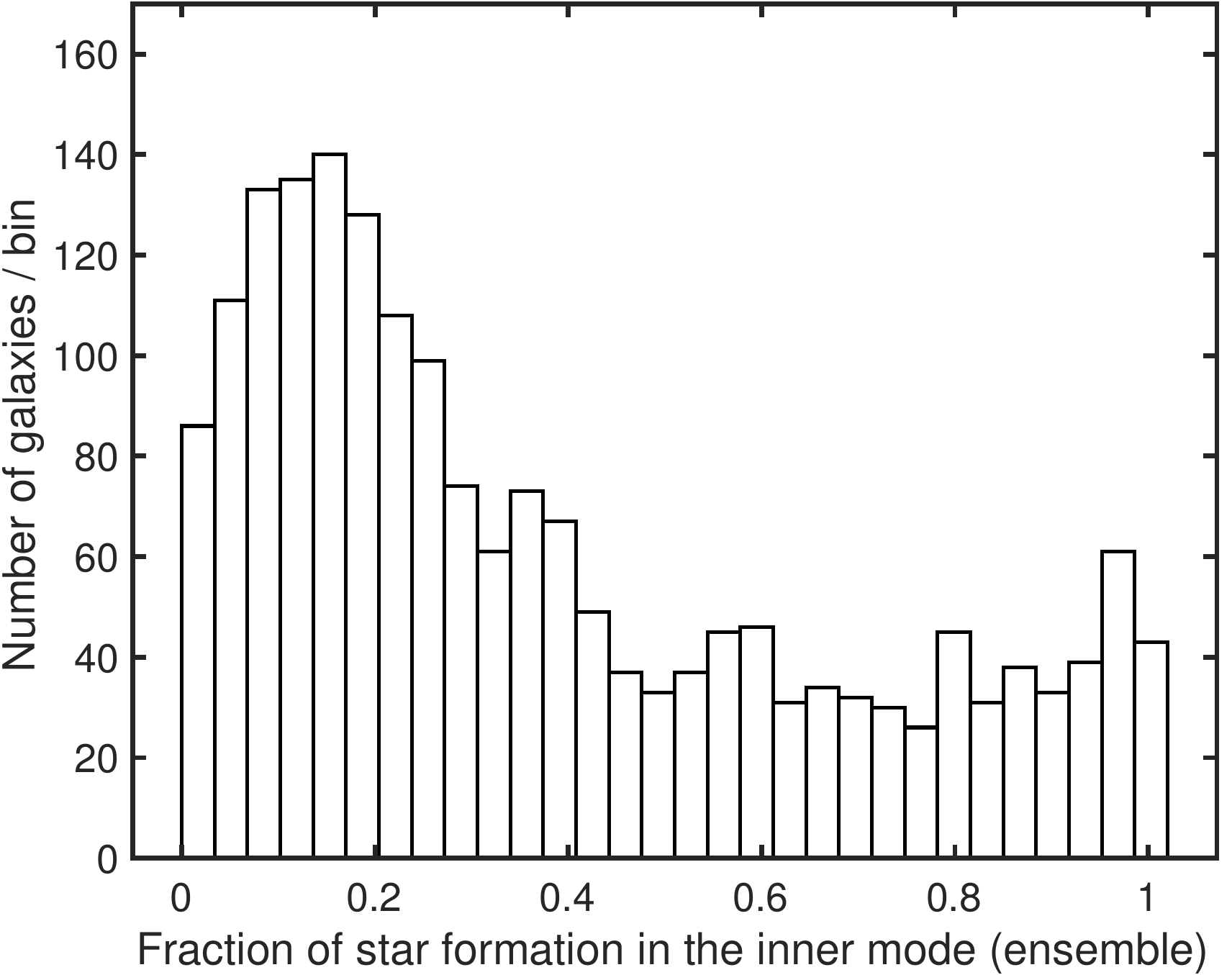}
\caption{Histogram depicting the fractional amount of star formation within individual TNG50 galaxies that falls within the inner mode relative to the outer mode of the $\Sigma_{\rm sfr}-$weighted ensemble $R$ distribution of the fiducial stellar mass bin at $z = 0$.}
\label{fig:fig8}
\end{figure}

As our next step, we seek to combine the two-component fit we obtained with the unweighted distribution (or the number distribution) of $R$ corresponding to the pixel sample (cf. log$R$ panel in Figure \ref{fig:fig2}). For this purpose, we obtain a functional approximation of the unweighted distribution by fitting it to a log-normal distribution of the form

\begin{equation}
    f_{\rm uw}(R) \equiv \frac{1}{N_{\rm tot}}\frac{dN(R)}{dR} = \frac{1}{R\sigma\sqrt{2\pi}} \exp{\frac{-({\rm ln}R - \mu)^2}{2\sigma^2}}
\end{equation}
Here, $N(R)$ denotes the number of pixels as a function of radius, $N_{\rm tot}$ is the total number of pixels, and $\mu$ and $\sigma$ are the mean and standard deviation of the distribution respectively.

Equipped with a functional form for both the average radial $\Sigma_{\rm sfr}$-profile as well the unweighted distribution, we then proceed to derive the resultant weighted distribution as

\begin{equation}
\begin{split}
     f_{\rm w}(R) & \equiv \frac{1}{\Sigma_{\rm sfr, tot}}\frac{d\Sigma_{\rm sfr}(R)}{dR}\\
     & = \frac{1}{\Sigma_{\rm sfr, tot}} \frac{d\Sigma_{\rm sfr}}{dN}(R) \frac{dN(R)}{dR}\\
     & = \langle\Sigma_{\rm sfr}(R)\rangle  f_{\rm uw}(R) \left(\frac{\Sigma_{\rm sfr, tot}}{N_{\rm tot}}\right)^{-1}
\end{split}
\end{equation}
where $\Sigma_{\rm sfr, tot}$ is the sum of $\Sigma_{\rm sfr}$ values of all pixels in the fiducial dataset.

As shown in Figure \ref{fig:fig5} bottom panel, the distribution so obtained not only reproduces the double-peaked structure of the original weighted distribution, but also, the two individual modes present can be separately recovered from the convolution of the unweighted distribution with the ``inner-" and ``outer-" component of the $\Sigma_{\rm sfr}(R)$-profile respectively. The multiplicity of modes in the weighted distribution is therefore a natural outcome of the occurrence of multiple exponential scale lengths in the corresponding average $\Sigma_{\rm sfr}(R)$-profile, with the scale length values also governing the positions of the modes.

\begin{figure}[t!]
\centering
\includegraphics[width=0.48\textwidth]{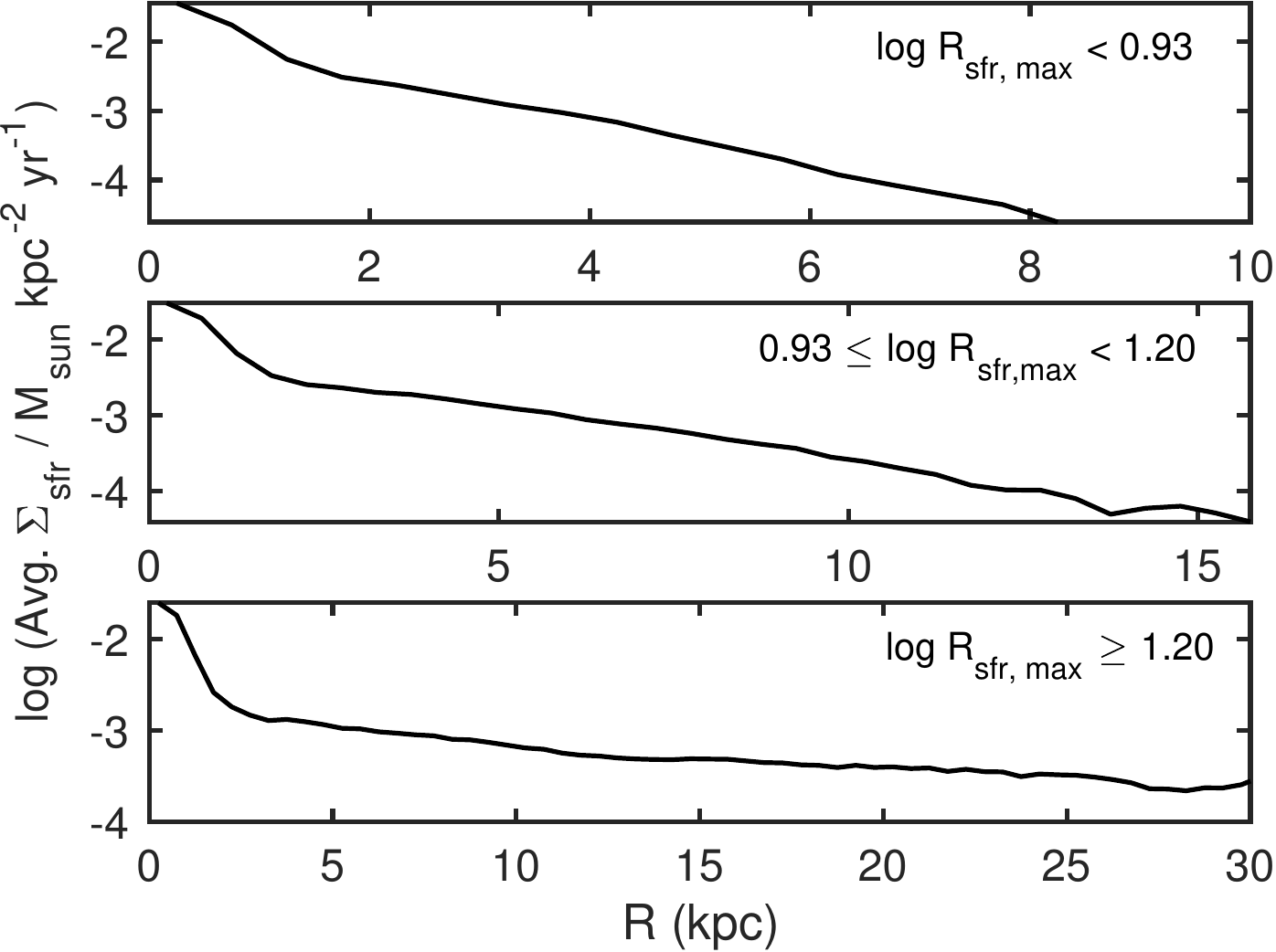}
\caption{Average radial star formation rate surface density profiles for the TNG50 fiducial galaxy sample separated into three different star-formation-size bins containing roughly equal number of galaxies.}
\label{fig:fig8-2}
\end{figure}

\begin{figure*}
\centering
    \includegraphics[width=0.95\textwidth]{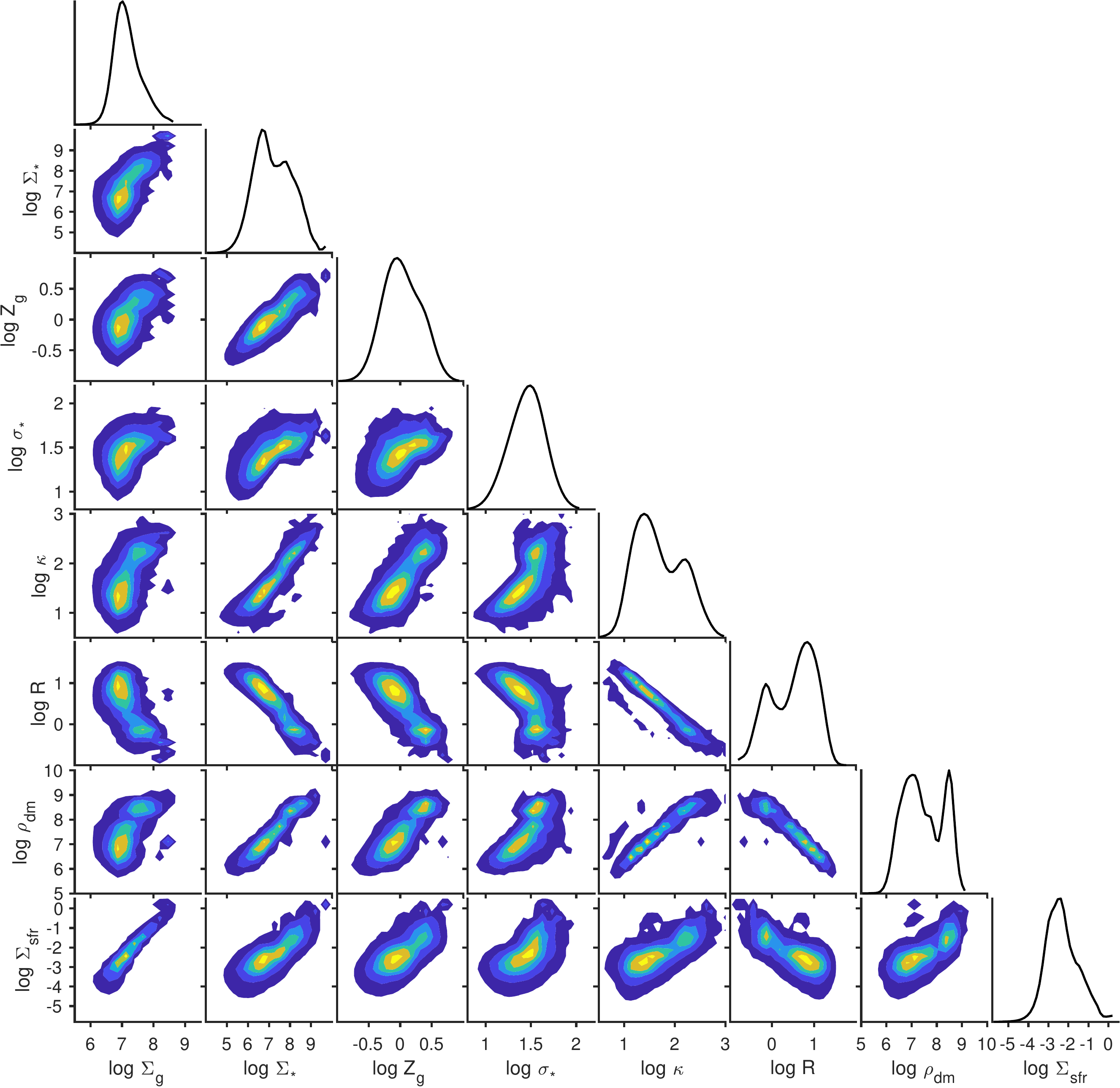}
\caption{Corner plot depicting scaling relations between properties of kiloparsec-sized star-forming regions in TNG50. Densities are derived using the collective data corresponding to all star-forming pixels from our fiducial sample. The diagonal panels show 1D $\Sigma_{\rm sfr}$-weighted histograms for each quantity, while coloured contours represent $\Sigma_{\rm sfr}$-weighted cumulative joint density fractions at 10, 30, 50, 70, 90 and 99\% from innermost to the outermost.} 
\label{fig:fig10}
\end{figure*}

\begin{figure*}[t!]
\centering
\includegraphics[width=\textwidth]{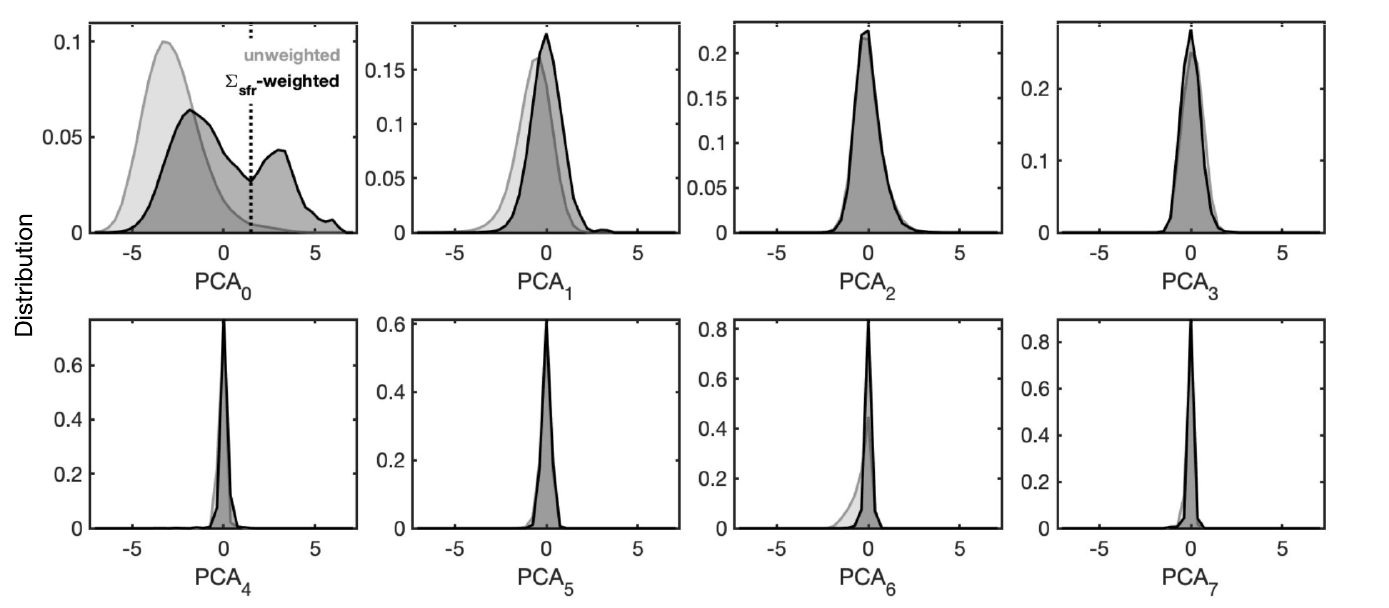}
\caption{The (unweighted)$\Sigma_{\rm sfr}$-weighted distributions of principal component scores in descending order of explained variance shown in grey(black). The first PC captures the bimodality present in multiple properties constituting the original parameter space. Due to the log-transform applied before conducting the PCA, the scores are unitless.}
\label{fig:fig11}
\end{figure*}

Having discerned the origins of bimodality in the ensemble property distributions, we now examine whether, and to what degree, the dichotomy in star formation conditions arises from two distinct populations of galaxies contributing exclusively to either peaks. In other words, could the bimodality result from an admixture of galaxies undergoing inside-out quenching producing the ``outer" mode with the ones with mostly central star formation manifesting as the ``inner" mode? Or perhaps a population of small galaxies with steep exponential profiles mixed with a separate population of large galaxies with shallow profiles? To this end, we assimilate the pixels corresponding to each galaxy separately and calculate the relative fraction of the galaxy's total star formation contained within the two modes of the ensemble distribution. To separate the two modes, we utilise the local minimum between the peaks as the separation radius, which for our fiducial sample turns out to be at log$R \simeq0.19$. We use this simplified criterion instead of a Gaussian mixture model (GMM) for peak separation as GMMs cannot be applied to density distributions when sample weights are in consideration. Figure \ref{fig:fig8} shows the distribution of the fractional amount of integrated star formation rate of each galaxy that is contained within the ``inner" mode (generated by the inner exponential component) of the weighted ensemble distribution of Figure \ref{fig:fig5}. As can be gleaned from the figure, the distribution spans the full range from 0 to 1 indicating that galaxies do not in fact fall into two separate classes, viz., star-forming cores and rings, contributing exclusively to either of the two modes of star formation at the ensemble level. While this is true, this does not deliver a clear conclusion about whether multiple components exist within the star-formation rate profiles of individual galaxies.

The existence of a sharp transition in the slope of the ensemble $\Sigma_{\rm sfr}$-profile may arise from two plausible scenarios: i) a population of small galaxies with steep profiles contributing to the inner exponential mixed with a population of large galaxies with shallow profiles producing the outer exponential, or ii) the preponderance of galaxies among the entire sample having the presence of two different scale lengths individually. To ascertain this, we investigate the shape of the average star-formation rate profile for our fiducial galaxy sample separated by their sizes as shown in Figure \ref{fig:fig8-2}. Galaxies are split into three equally populated bins of \emph{star formation size}, which we define to be the galactocentric radius of the farthest star-forming pixel in each galaxy. The panels confirm that galaxies of different sizes all have the presence of a broken star formation rate profile, albeit with a more pronounced transition (or elbow) in bigger galaxies on account of their outer components having greater scale lengths, and hence, shallower slopes relative to the inner component. Combined with the inference from the previous figure, this finding suggests that galaxies in general provide a finite contribution to the two modes of star-formation that are present in the ensemble distribution. Additionally, the overall skewness of the distribution in Figure \ref{fig:fig8} towards lower values implies that the vast majority of galaxies belonging to the fiducial sample exhibit a greater amount of star formation in their diffuse outskirts also at the individual galaxy level, which is analogous to and confirms the population-wide trend noted in \S\ref{sec4}.

\section{A multi-dimensional view of the ISM parameter space}\label{sec5}
\subsection{Resolved Galaxy Scaling Relations}\label{sec5.1}

Having thus far examined and discussed the statistical nature of our physical parameter space one quantity at a time, we now look into the mutual relationships amongst these spatially-resolved ISM properties in a pairwise fashion, more commonly known as \emph{resolved scaling relationships}, predicted by TNG50.

In Figure \ref{fig:fig10}, we present the joint distribution functions of all possible pairs of physical properties measured for star-forming regions belonging to the galaxies in our fiducial sample. The diagonal panels show one-dimensional $\Sigma_{\rm sfr}$-weighted probability density distributions for the property labeled on the corresponding abscissa (same as in Figure \ref{fig:fig2}), while each of the off-diagonal panels show the \emph{$\Sigma_{\rm sfr}$-weighted} two-dimensional cumulative density contours (upto 99\% represented by the outermost contour) for the property pair indicated by the corresponding axes labels. In the following discussion, our use of the term \emph{linear} is meant to indicate linearity in log-space.

Our results give rise to spatially-resolved counterparts to several canonical global galaxy scaling relationships  \citep{tremonti04,noeske07,schmidt59, kennicutt98}, namely, the mass-metallicity relation ($\Sigma_{\star}-Z_{\rm g}$) \citep{sanchez19, sanchez17, barrera16}, star formation main sequence ($\Sigma_{\star}-\Sigma_{\rm sfr}$) \citep{abdurrouf18, liu18, cano-diaz16} and the Schmidt star-formation law ($\Sigma_{\rm g}-\Sigma_{\rm sfr}$) \citep{calzetti18, roychowdhury15, leroy13, schruba10, bigiel08}. Apart from these, the figure demonstrates a widespread presence of linear or near-linear relationships between multiple other quantities. The existence of some correlations such as those with galactocentric radius is intuitively expected on account of structural and dynamical considerations as most galaxies are known to have well-defined density and metallicity profiles. However, other correlations (for e.g. metallicity vs. stellar velocity dispersion) have seemingly less obvious physical origins and warrant detailed investigation in a separate future study (P. Torrey et al., in preparation).

We notice that some relationships, specifically the Schmidt law and mutual relations between stellar mass density, dark-matter density, radius and the epicyclic frequency are extremely tight with negligible scatter. In the case of the former, the low scatter indicates that in TNG50, star formation is not only very closely related to the mass of the gas \citepalias[due to][]{sh03}, but is also well-sampled on kpc-scales across the entire range of $\Sigma_{\rm g}$. The latter set of relationships simply reflect the empirically known $\Sigma_\star$ and $\rho_{\rm dm}$ radial profiles, as well as the definition used for $\kappa$ in our analysis (\S\ref{sec2}). Contrary to this, relationships involving gas density (with the exception of the Schmidt law) exhibit little to no correlation. These characteristics also come up in our subsequent analysis in \S\ref{sec5.2}.

In a similar vein of contrasting features, we find that while a majority of the scaling relationships have a monotonic and manifestly linear shape, others (most notably, all panels representing stellar velocity dispersion) hint at a break or a turnover. Finally, in many of the two-dimensional distributions, the underlying bimodality in the physical quantities is manifested as two distinct clouds, which in some cases deviate from one another in terms of their slope, thereby giving rise to the aforementioned broken scaling relationships.

The ubiquity of linear correlations in Figure \ref{fig:fig10} points to a high degree of redundancy in the overall parameter space, where multiple parameters encode shared information amongst them and lessen the effective degrees of freedom or ``axes of variance" available. This information, in principle, should therefore be accessible using a lower-dimensional representation of the same space. Motivated by this feature, we subsequently embark on a search for a reduced representation of the ISM hyper-parameter space using the commonly used technique of principal component analysis for dimensionality-reduction.

\begin{deluxetable*}{l>{\columncolor[gray]{0.9}}c>{\columncolor[gray]{0.9}}c>{\columncolor[gray]{0.9}}c>{\columncolor[gray]{0.9}}c>{\columncolor[gray]{0.9}}c>{\columncolor[gray]{0.9}}c>{\columncolor[gray]{0.9}}c>{\columncolor[gray]{0.9}}cc}
% \begin{deluxetable*}{l|cccccccc|c}
\tablenum{2}
\label{table:pca}
\tablecaption{Results of the weighted principal component analysis on the full dataset corresponding to the fiducial galaxy population. Each row describes the linear coefficients (loadings) associated with the 8 physical parameters (features) that make-up the corresponding principal component. \label{tab:tab3}}
\tablewidth{0pt}
\tablehead{
\colhead{Component} & \colhead{$\Sigma_{\rm g}$} & \colhead{$\Sigma_{\star}$} & \colhead{$Z_{\rm g}$} & \colhead{$\Sigma_{\rm sfr}$} & \colhead{$\sigma_{\star}$} & \colhead{$\kappa$} & \colhead{$R$} & \colhead{$\rho_{\rm dm}$} & \colhead{\% variance explained} 
}
\startdata
PC$_0$ & 0.3147 &    0.3867 &    0.3437 &    0.3526 &    0.2968 &    0.3803 &   -0.3769 &    0.3664 & 78.83\\
PC$_1$ & 0.7034 &   -0.1131 &   -0.1561 &    0.5209 &   -0.1869 &   -0.1757 &    0.1961 &   -0.3041 & 9.13 \\
PC$_2$ & 0.0362 &   -0.0771 &   -0.3897 &    0.0114 &    0.8965 &   -0.0920 &    0.1558 &   -0.0656 & 6.29 \\
PC$_3$ & 0.0475 &   -0.1157 &   -0.7803 &    0.0088 &   -0.2322 &    0.2838 &   -0.3937 &    0.2933 & 3.20 \\
PC$_4$ & 0.1226 &   -0.0775 &    0.0169 &    0.0011 &   -0.0414 &   -0.5489 &    0.2223 &    0.7914 & 1.20 \\
PC$_5$ & -0.0707 &   0.8617 &   -0.3048 &   -0.0204 &   -0.1254 &   -0.0250 &    0.3770 &   -0.0279 & 0.65 \\
PC$_6$ & -0.6120 &  -0.0209 &   -0.0522 &    0.7605 &   -0.0149 &   -0.1900 &   -0.0868 &   -0.0153 & 0.36 \\
PC$_7$ & -0.0903 &  -0.2635 &    0.0306 &    0.1594 &   -0.0412 &    0.6305 &    0.6642 &    0.2359 & 0.34 \\
\enddata
\tablecomments{The 8 x 8 matrix of shaded values constitutes the transpose of the coefficient matrix ({\bf C$^{\rm T}$}; with rows corresponding to PCs and columns corresponding to features), and can be used for the reconstruction of the original parameter space from the principal component values (ref. Appendix \ref{appendixB} for the exact procedure).}
\end{deluxetable*}

% PC$_0$ & 0.3347 &   0.4130 &   0.4473  &  0.3639 &   0.2542 &   0.3787 &  -0.3971 &   0.3724 & 78.83\\
% PC$_1$ & 0.7481 &  -0.1208 &  -0.2031  &  0.5376 &  -0.1601 &  -0.1750 &   0.2066 &  -0.3091 & 9.13\\
% PC$_2$ & 0.0385 &  -0.0823 &  -0.5072  &  0.0117 &   0.7678 &  -0.0916 &   0.1642 &  -0.0666 & 6.29\\
% PC$_3$ & 0.0505 &  -0.1236 &  -1.0155  &  0.0091 &  -0.1989 &   0.2827 &  -0.4148 &   0.2981 & 3.2\\
% PC$_4$ & 0.1304 &  -0.0827 &   0.0220  &  0.0011 &  -0.0354 &  -0.5466 &   0.2343 &   0.8044 & 1.2\\
% PC$_5$ & -0.0752 &  0.9202 &  -0.3967  & -0.0210 &  -0.1074 &  -0.0249 &   0.3972 &  -0.0284 & 0.65\\
% PC$_6$ & -0.6509 & -0.0223 &  -0.0679  &  0.7848 &  -0.0128 &  -0.1892 &  -0.0915 &  -0.0156 & 0.36\\
% PC$_7$ & -0.0961 & -0.2814 &   0.0398  &  0.1645 &  -0.0353 &   0.6279 &   0.6999 &   0.2397 & 0.34\\

\subsection{Characterising the Hyperspace of ISM Properties}\label{sec5.2}
\label{sec:pca}

Even though each star-forming region in our analysis is represented by a set of multiple physical parameters, not all of them are expected to be equally informative. Sometimes, relationships between parameters exist (as evident in \S\ref{sec5.1}) thereby lowering the degrees of freedom needed to account for the information contained in the original space, also known as the intrinsic dimensionality of the dataset. In order to better scrutinise what these relationships are, and their implications for the conditions in which star formation takes place, we now conduct a statistical characterisation of our measured 8D parameter space by means of lowering its dimensionality using the technique of principal component analysis (PCA). Through this exercise, we seek to answer the question: Is there a simplified meaningful representation of the underlying distribution of star formation in the ISM, and if yes, how many controlling parameters are required for such a representation to work?

\subsubsection{Principal Component Analysis}\label{sec5.2.1}

Principal component analysis is a widely-used non-parametric analysis technique to reveal low-dimensional representations of structures underlying complex high-dimensional datasets. It does so by quantifying the importance of each original dimension for describing the information content (or variance) contained within the data. Mathematically, PCA aims to re-express a given dataset with a new set of orthogonal basis - constructed through a \emph{linear combination} of its original basis - that minimises noise and whose axes (known as principal components or PCs) preserve most of the variance within the dataset. This essentially amounts to the diagonalisation of the covariance matrix of the dataset (which carries information about the redundancy between parameters and overall noise in the data) using eigenvalue decomposition. The PCs thus obtained are naturally uncorrelated, and are traditionally expressed as a rank-ordered set based on their corresponding variances. Accordingly, the leading component PC$_0$ is aligned with the direction of largest variance in the data, followed by PC$_1$, PC$_2$ and so forth. A lower (say, $k$) dimensional \emph{hyperplane} approximation to the initial space is then achieved by defining a threshold variance and keeping only the first $k$ PCs required to capture that amount of variance.

A common practice in the application of PCA is to transform and standardise the data to bring all variables on the same footing in terms of their magnitude and scales. Given the vastly different units of measurements and dynamic ranges associated with ISM properties within our dataset, we apply a logarithmic transformation to our entire dataset. In doing so, we alleviate the impact of skewness of the distributions in the linear space, and attain a more practically useful sampling resolution across the dynamic range of all the properties. In addition to this transformation, we also standardise our data such that each original dimension is re-scaled to have a mean of zero and unit variance. This is done to avoid variables with larger scales from dominating the covariance structure of the dataset and biasing the directions of the PCs. Taking this data standardisation step then makes the PCA procedure equivalent to diagonalising the correlation matrix instead of the covariance matrix.

\begin{figure}[t]
\centering
\includegraphics[width=0.47\textwidth]{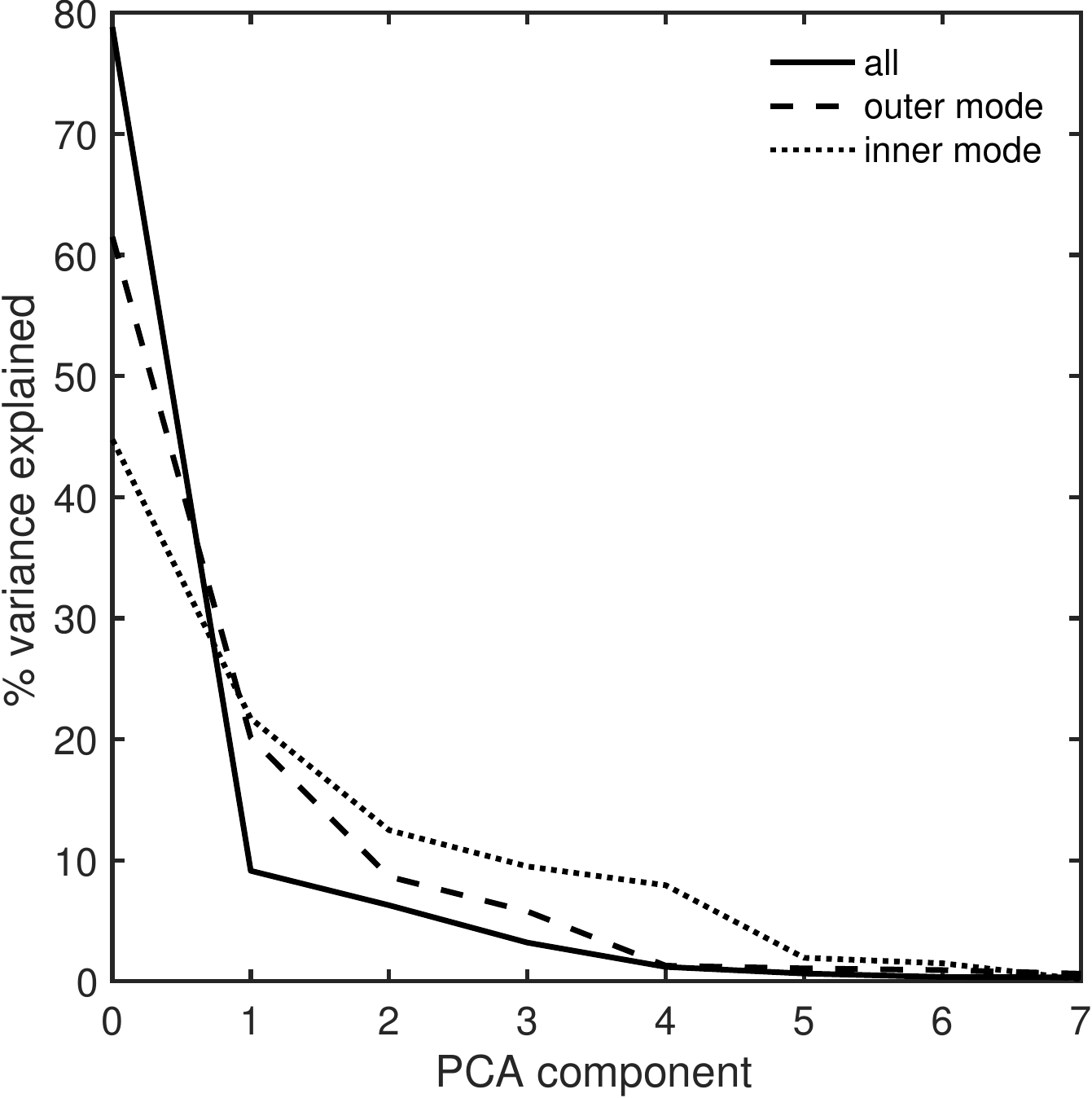}
\caption{Percentage of variance in the data explained by each of the principal components.}
\label{fig:fig12}
\end{figure}

\begin{figure*}
\centering
\includegraphics[width=0.9\textwidth]{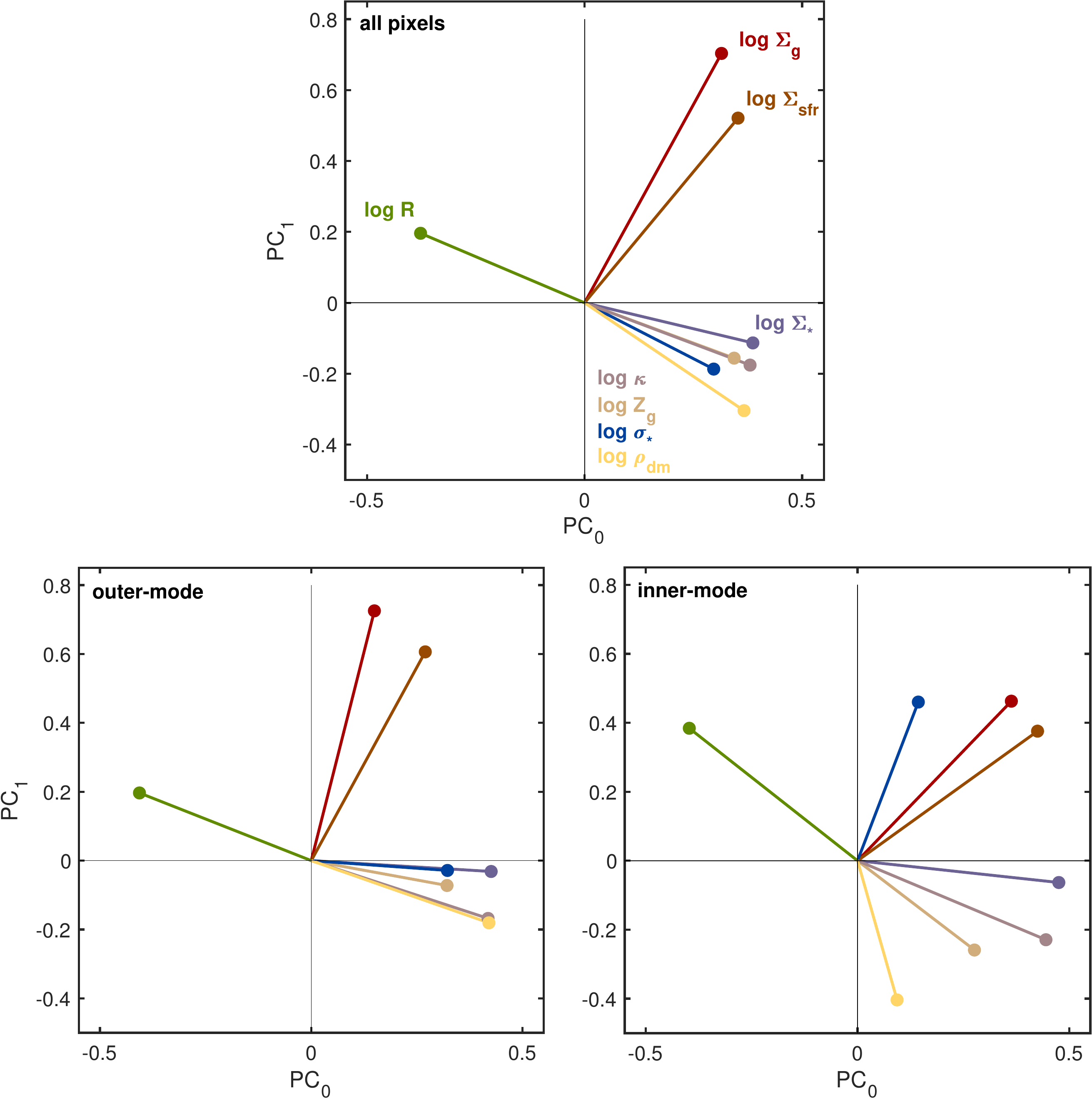}
\caption{Loading plots from the $\Sigma_{\rm sfr}$-weighted principal component analysis for pixels corresponding to our fiducial sample. Coloured lines with markers indicate the coefficients of various physical quantities associated with the first and second principal components for the entire sample (top), the outer-mode (bottom-left), and the inner-mode (bottom-right).}
\label{fig:fig13}
\end{figure*}

In our study, we use the correlation-matrix-based PCA implementation available in \texttt{MATLAB}, which uses a singular-value decomposition algorithm in lieu of the more traditional eigenvalue decomposition to compute the principal component axes. Through the rest of this section, we will refer to the individual star-forming pixels as \emph{samples}, and the corresponding ISM properties as \emph{features}. The dataset derived from our fiducial sample of galaxies and analysed herein constitutes a total of 589,822 samples each associated with an 8-dimensional feature vector.

\subsubsection{PCA Results}\label{sec5.2.2}

The principal components we get from our analysis represent a new set of axes (obtained by a generalised rotation of the initial basis) onto which the original variables can be projected. By virtue of being a simple transformation of the original coordinates, the principal component space is expected to capture the most important characteristics (such as linear correlations) of the feature space. It is therefore instructive to look at the distributions of these transformed coordinates, known as component ``scores", shown here in Figure \ref{fig:fig11}. As evident from the figure, the weighted distribution of the leading component PC$_0$ has a bimodal shape while the other PCs have distributions unimodal in nature. This indicates that PC$_0$ by itself is able to capture the distinction between the two star formation regimes exhibited in multiple dimensions in the original feature space. Taking advantage of this fact, we proceed to split our full data into two separate datasets corresponding to the aforementioned regimes (cf. \S\ref{sec4}) by making a cut along the PC$_0$ axis, which is determined by the point of minimum between the two peaks. We conduct PCA separately on these two subsets of the dataset so as to allow us to better understand which physical correlations govern the dispersion within each of the two regimes. Hereafter, the `outer-mode' stands for the low-value peak of PC$_0$ corresponding to the high-radius low-density star-forming region population, whereas the `inner-mode' represents the low-radius (central) high-density population.

In Figure \ref{fig:fig12}, we show the percentage of the total variance explained by each PC obtained from the decomposition of the entire fiducial dataset, and the two modes separately. For the combined data, the first component (PC$_0$) alone captures $\sim$80\% of the total variance in the dataset, the second one (PC$_1$) $\sim$9\%, the third (PC$_2$) $\sim$6\%, and the values drop sharply thereafter with the last few PCs presumably representing noise. The widths of the distributions in Figure \ref{fig:fig11} also portray this trend. That the first two PCs collectively account for $\sim$90\% of the total variability means that a 2D projection of the feature space could indeed provide a reasonable characterisation of the complete 8D dataset. Furthermore, the bimodal shape of the PC$_0$ distribution alongside its high value of associated variance suggests that most variability in the data is manifested as peak-to-peak variance arising from the mix of two different sample populations (i.e., bimodality) in the dataset. In fact, this variance far exceeds the sample-to-sample variability within the peaks themselves. On the other hand, in case of the separated modes, the explained variance curve declines rather gradually and the number of PCs needed to capture 90\% of the variance goes from 2 up to 5, with the respective first components explaining far less variance (60 and 45\%) than their counterpart for the combined data.

Next, we turn towards quantitatively examining the relationship between the feature space and the resulting principal component space. In other words, since PCs are a linear combination of the ISM properties, we can determine the contribution of each of those properties to the PCs, also known as ``loading factors" (see Table \ref{table:pca}). Figure \ref{fig:fig13} depicts the eight ISM properties as 2D vectors plotted in the PC$_0$ (abscissa) vs. PC$_1$ (ordinate) plane for the full data (upper panel) as well as the two modes separately (lower panels). The (x, y) coordinates of each property are their loading factors along the corresponding PC direction. Several key attributes emerge:

\begin{figure}[t]
\centering
\includegraphics[width=0.47\textwidth]{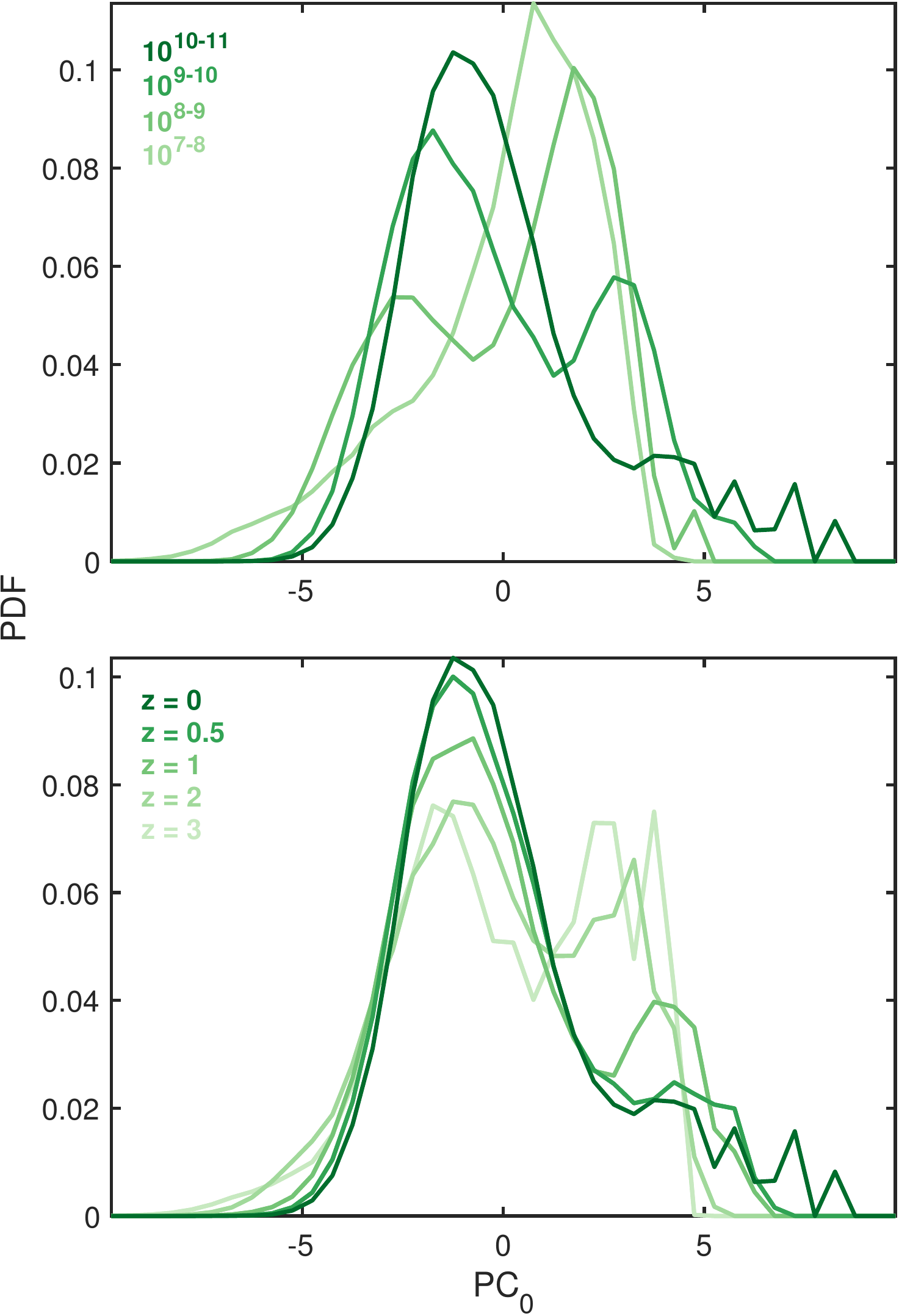}
\caption{Evolution of the probability density distribution of the first principal component with parent galaxy stellar mass (top) and redshift (bottom). All curves correspond to galaxies at $z = 0$ in the top panel, and the mass bin 10$^{10-11}$ M$_\odot$ in the bottom panel (consistent with Figures \ref{fig:fig3} and \ref{fig:fig4} respectively).}
\label{fig:fig14}
\end{figure}

\begin{figure*}
\centering
\includegraphics[width=\textwidth]{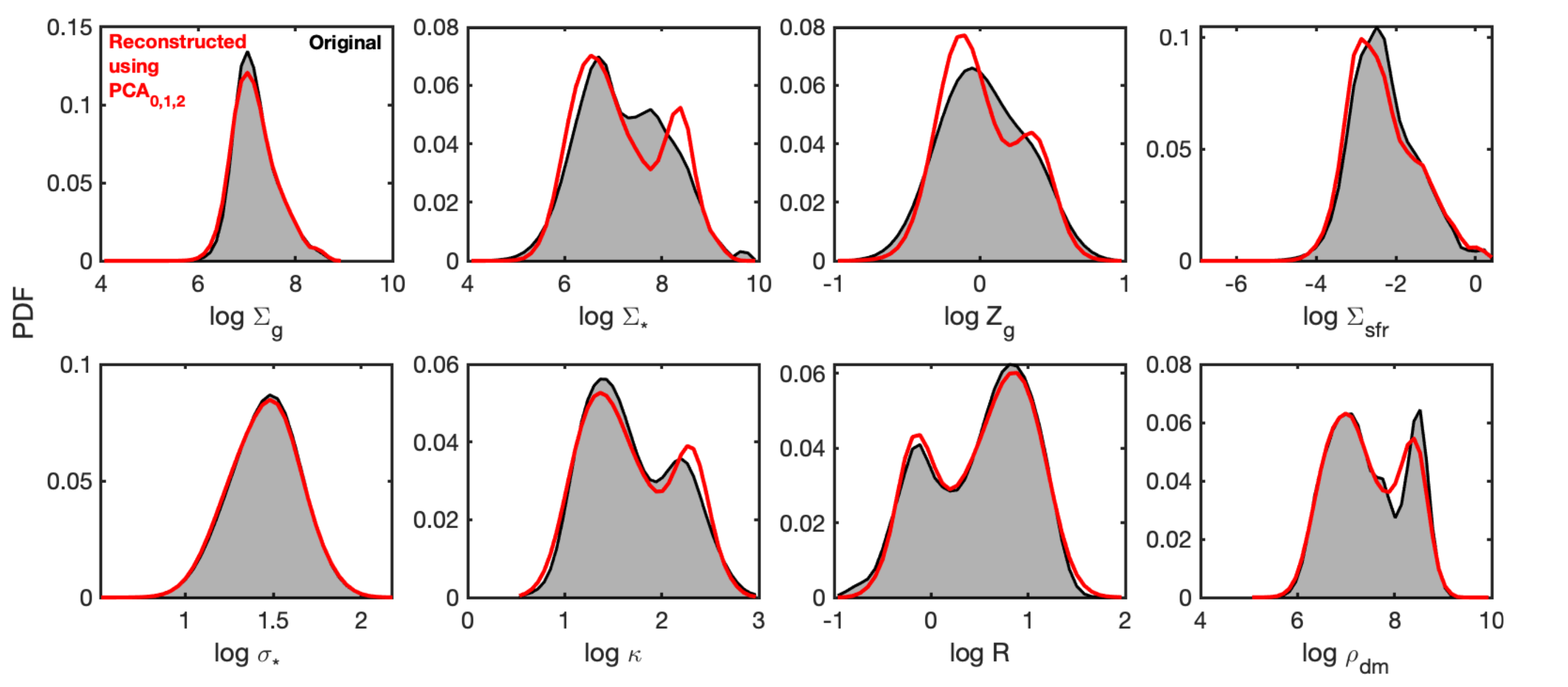}
\caption{A reconstruction of the original 8D physical parameter space using the first 3 principal components. See Appendix \ref{appendixB} for details on the exact reconstruction procedure.}
\label{fig:fig15}
\end{figure*}

From the top panel, we see that all properties contribute with roughly equal weights to the first principal component and, with the exception of galactocentric radius, have a similarly positive sign. This conveys a correlated equal variation of all quantities with this component to first order. Additionally, it explains why PC$_0$ of the full dataset captures the bimodality - at low values of PC$_0$, samples would have high values of $R$ and low values of all the other quantities thereby belonging to the high-R/low density regime i.e., the outer mode. By the same token, samples that have a high value of PC$_0$ would belong to the inner mode. A similar trend is seen in the PC$_0$ loadings associated with the outer mode, albeit with a reduced contribution of $\Sigma_{\rm g}$ and $\Sigma_{\rm sfr}$. Thus, PC$_0$ in this case highlights processes that modulate the environment within stellar discs, which is primarily governed by the dynamical influence of stars and dark-matter. In the inner mode, the pattern is significantly different and PC$_0$ loses most of its correlation with $\sigma_{\star}$ and $\rho_{\rm dm}$, hence tracing a more complex dynamical environment driven mainly by gas- and stellar-gravity. 

In addition to the composition of PCs, we can draw insights from Figure \ref{fig:fig13} pertaining to the correlations existing between the features themselves. In the top panel, we observe a clear clustering amongst ISM properties hinting at a high degree of multi-collinearity in the system. In particular, the quantities \{$\Sigma_{\star}, \kappa, Z_{\rm g}, \sigma_{\star}, \rho_{\rm dm}$\} lie in a tight cluster indicating that they have a strong positive correlation\footnote{the strength of the correlation goes as cos($\theta$), where $\theta$ is the angle between the vectors. This means that $\theta$ = 0(180$^\circ$) depicts perfect linear correlation(anti-correlation).} amongst themselves, and a strong negative correlation with the galactocentric radius $R$. $\Sigma_{\rm g}$ and $\Sigma_{\rm sfr}$ also have a positive correlation between them, which is an expected result based on the canonical Schmidt law. These observations are well in line with our interpretation of Figure \ref{fig:fig10}. In the case of $\Sigma_{\rm g}$, the apparent overall lack of linear correlation with other parameters noted in the previous subsection also emerges in the results of our PCA analysis. This observation is perhaps suggestive of the fact that the variance associated with the $\Sigma_{\rm g}$ scaling laws either comes from predominantly non-linear dependencies, or is identified as noise and hence not captured by the high-ranking PCs shown in Figure \ref{fig:fig12}. The figure also affords us the additional clarity that this behaviour is almost entirely on account of the more dominant \emph{outer} mode. In the case of the outer mode - which we previously saw to be representative of a stellar disc environment - the same correlations stand as in the case of the full dataset. However, in the \emph{inner} mode, most correlations barring the $R-\kappa-Z_{\rm g}$ and $\Sigma_{\rm g} - \Sigma_{\rm sfr}$ relationships are appreciably weakened. Strikingly, in these central star-forming regions, $\sigma_{\star}$ is no longer linked with the properties of stars but is more tightly coupled to the gas instead. Furthermore, $\sigma_\star$ does not show a strong correlation with the radius, signalling the near-flatness or lack of a well-defined vertical velocity-dispersion profile in those regions.

It is worth keeping in mind here that PCA, being a linear method, is not expected to fully capture the non-linear aspects of relationships between features, if present. Given that, in order to describe a non-linear relationship fully, one cannot rely on a single principal component. Rather, in such a case, a group of PCs is needed, of which one would provide the best linear-approximation to the underlying relationship while the others would encompass variances in the directions of deviations to non-linearity. By using a logarithmic transform, however, we are not strictly confined to the linear regime and are able to additionally capture power-law relationships. Although non-linear generalisations of PCA (such as autoencoder neural networks and kernel-PCA) as well as more advanced manifold-learning methods exist, they are not conducive to the kind of analysis we have conducted in this study i.e., one that requires taking individual sample weights into account.

While so far we stepped through the PCA analysis results for our fiducial sample of galaxies, we now briefly consider how these results change when we conduct PCA on samples representing different galaxy stellar mass ranges and redshifts (akin to our approach in \S\ref{sec4.2}, \ref{sec4.3}). In Figure \ref{fig:fig14}, we show the distribution of PC$_0$ for star-forming region datasets derived from galaxies in different stellar mass bins at $z$ = 0, and for datasets corresponding to galaxies with M$_\star$ = 10$^{10-11}$ M$_\odot$ at different redshifts. Darker curves represent higher stellar masses and lower redshifts. The variation in the relative amplitude of the two peaks as a function of M$_{\star}$ is reproduced well (compared to those observed in \S\ref{sec4.2} and \ref{sec4.3} for the full space), as is the trend with redshift, where the preference for star formation gradually shifts towards low density, disc-like extended environments for lower redshifts and higher stellar masses. This finding reinforces our understanding from previous results that the leading component fully captures the bimodality signature in the original features as well as the inherent physical relationships that correlate them.

Lastly, in Figure \ref{fig:fig15}, we compare the actual distributions of the initial physical space parameters for our fiducial sample of galaxies with the reconstructed versions generated only by including the first three principal components. We find that the agreement between the original and the low-dimensional version is excellent. For all the physical parameters where there are clearly two distinct peaks, we recover their respective locations and the position of the minimum between them. On the other hand, in terms of the relative peak heights between the two modes, the reconstructed representations are somewhat discrepant from the original distributions. As such, it is possible to reproduce the presence of bimodality in quantity distributions by using PC$_0$ alone, however, capturing their exact attributes and meaningful variations within individual peaks requires additional components. In our case, the first three PCs account for $\sim$94\% of the information present in the dataset (see Table \ref{table:pca}). Hence, by retaining a substantial fraction of the information, a three-dimensional embedding of our dataset can serve as a practically useful avenue for sampling a family of star-forming regions for further specific investigations, such as tall-box simulations to study feedback and gas dynamics. As a supplementary data product, we provide the PCA loading factors, variances explained by all of the PCs, and joint distributions of the first three PC scores for star-forming regions from galaxies with M$_\star = 10^{7-8}, 10^{8-9}, 10^{9-10}, 10^{10-11}$ M$_\odot$ at $z = \{0, 0.5, 1, 2, 3\}$ at \url{https://github.com/bhawnamotwani/smaug}.

\section{Preliminary Comparison with Observations}\label{sec6}

To assess some of the results obtained from our simulations in the context of observations, we now proceed to examine the distributions of properties from resolved observations of nearby star-forming galaxies from the MaNGA IFU survey data \citep{bundy15}. Given the resolution of TNG50 and the scale chosen for our analysis, MaNGA offers a suitable dataset for a qualitative comparison against our results. However, due to the unavailability of an observational counterpart to several of the `local' properties we have worked with, we limit the comparison in this section only to a handful of quantities, namely, $\Sigma_{\star}$, $\Sigma_{\rm sfr}$, and $R$.

\subsection{Survey Description}\label{sec6.1}

The MaNGA survey is one of the three programs undertaken as part of the fourth installment of the Sloan Digital Sky Survey (SDSS-IV) aimed at observing resolved kinematic structure of $\sim$10,000 nearby galaxies through integral field spectroscopy. The imaging and spectroscopy for the galaxies is conducted using the 2.5m telescope at the Apache Point Observatory (APO) alongside specialised IFUs and the BOSS spectrograph with coverage in the 3600-10300A range and resolving power R $\sim 2000$. In this work, we use the latest public release DR15 \citep{aguado19} consisting of 3D data-cubes for a sample of 4824 unique galaxies uniformly sampled over the stellar mass range $\sim10^{9-11}\ {\rm M}_\odot$.

Raw data is reduced using the MaNGA's internal data reduction pipeline \citep{law16} and analyzed using the data analysis pipeline \citep{westfall19, belfiore19}. The local mass density of spaxels used in this work is computed as part of the Pipe3D pipeline through stellar population modelling by performing a linear decomposition of the spectrum into simple stellar populations of different ages and metallicities for each spaxel and correcting for dust attenuation (using the Balmer decrement) prior to fitting \citep[for full details, see][]{sanchez16}. The local star formation rates are derived from H$\alpha$ luminosity using the formula from \citet{kennicutt98} and a \citet{chabrier03} initial mass function. Due to an imposed threshold of S/N = 3 on H$\beta$ (used for extinction-correction) alongside distance and intrinsic luminosity constraints, the effective median sensitivity limit of $\Sigma_{\rm sfr}$ in MaNGA lies at $\sim10^{-3}$ M$_\odot$ ${\rm yr}^{-1}$ ${\rm kpc}^{-2}$. The spatial coverage of galaxies is expected to be at minimum upto 1.5 effective radii ($R_{\rm e}$). Lastly, MaNGA observations have a median spatial resolution of 2.5" FWHM ($\simeq$ 1.8 kpc at the median redshift of $\approx$ 0.03), and are sampled at a scale of 0.5" per spaxel in the final data cubes which translates into a physical scale of $\sim$1-2 kpc per spaxel given the redshift range of 0.01 $<$ z $<$ 0.15.

\begin{figure}[t!]
\centering
\includegraphics[width=0.48\textwidth]{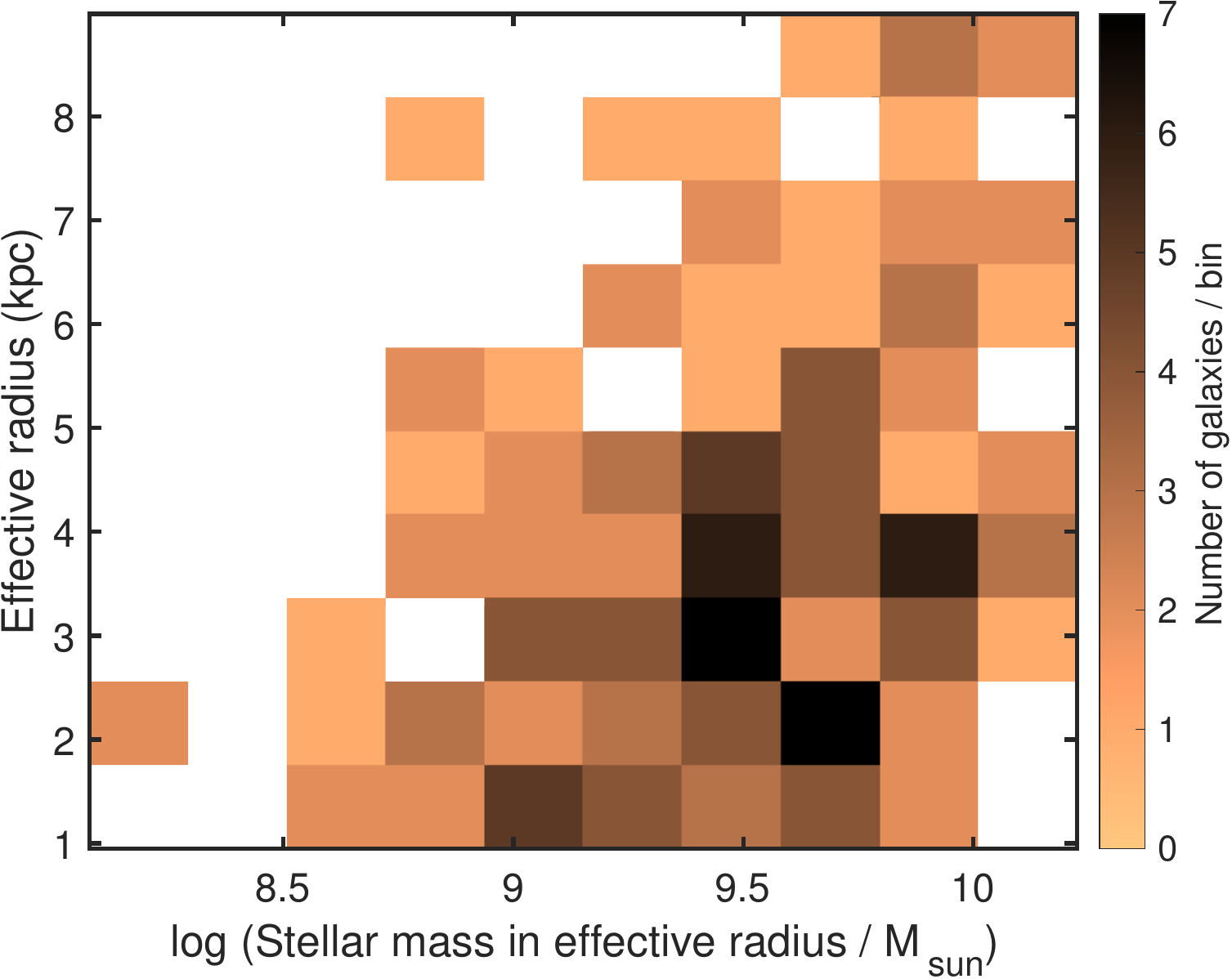}
\caption{The size-mass distribution of TNG50 and MaNGA galaxy samples used for comparative analysis in this section. The size and mass definitions are as described in the axes labels (with the size defined in a 2D face-on projection), and detailed further in the text.}
\label{fig:fig7}
\end{figure}

\begin{figure*}[t!]
\centering
\includegraphics[width=\textwidth]{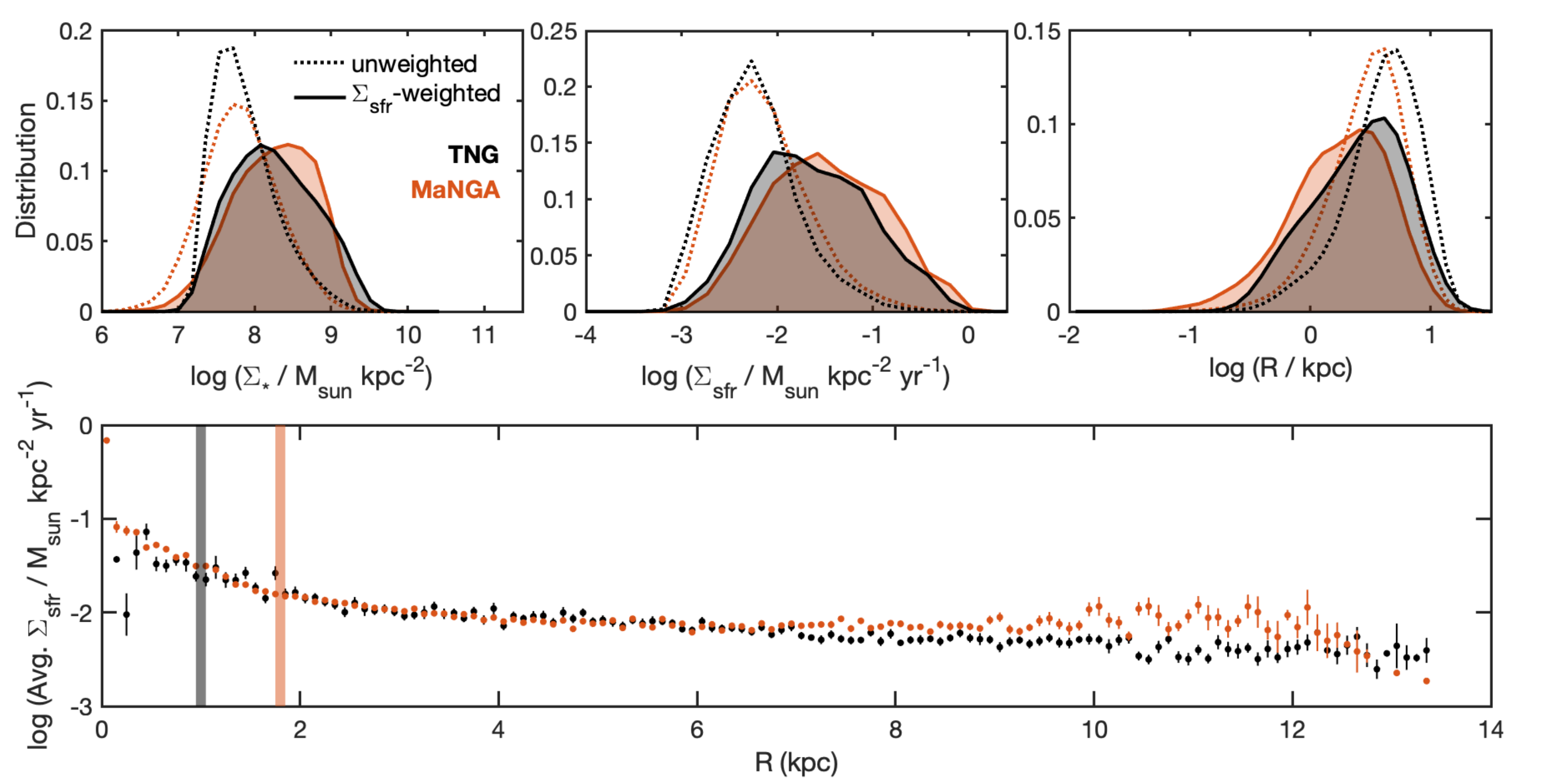}
\caption{Top: Probability density distributions for $\Sigma_{\star}$ (left), $\Sigma_{\rm sfr}$ (middle) and $R$ (right) for all well-defined star-forming spaxels in MaNGA (orange) at $z = 0-0.02$ compared with those from TNG50 (black) at $z = 0$, derived in both cases from galaxies with total stellar mass tentatively in the [$10^9$, $10^{10}$] M$_\odot$ range. Dotted lines represent the intial unweighted distributions while the solid curves are weighted by $\Sigma_{\rm sfr}$. Bottom: Binned average radial star formation rate surface density profiles for MaNGA and TNG50, with the errorbars indicating relative error of mean in each bin. Vertical bars in grey and orange denote the analysis scale used for TNG50 (1 kpc) and the spatial resolution of MaNGA at median redshift ($\simeq$ 1.8 kpc) respectively.}
\label{fig:fig9}
\end{figure*}

\subsection{Galaxy and Spaxel Selection}\label{sec6.2}

For our analysis, we choose all galaxies from MaNGA DR15 with a given BPT classification \citep{kewley01, kauffmann03} of either `cLIER' (galaxies with kpc-scale low-ionisation emission regions in their centres accompanied by star-formation in the outskirts) or `star-forming' \citep[as determined by][]{belfiore18}. To select a comparison sample for both MaNGA and TNG, we implement a selection criterion that matches the simuated and observed galaxies in their size-mass plane by stochastically sampling them in a binned fashion as shown in Figure \ref{fig:fig7}. Specifically, we use a metric for the two-dimensional effective radius of the galaxy and the total stellar mass enclosed within the effective radius calculated from the corresponding pixels or spaxels for each galaxy. For MaNGA, we adopt the effective radius to be the inclination-corrected Petrosian half-light radius ($R_e$), and the face-on projected 2D \emph{r}-band half-light radius ($R_{1/2,\rm r-band}$) for TNG50 galaxies. The particular choice of mass inside $R_{1/2,\rm r-band}$ for TNG50 is made in order to avoid uncertainties arising from the differences between the definitions of total mass in the simulation (all mass within the 3D virial radius) and in observation (mass calculated from light within twice the 2D Petrosian radius). Following this procedure gives us a total of only 147 galaxies in both MaNGA and TNG due to the highly non-overlapping nature of their initial distributions. Thereafter, for all the selected galaxies, we convolve our images with a MaNGA-like Gaussian PSF with FWHM$\simeq$1.8 kpc, and then draw the corresponding spaxel contributions from within 1.5 times $R_e$ ($R_{1/2,\rm r-band}$) in MaNGA(TNG) so as to record their $\Sigma_{\star}, \Sigma_{\rm sfr}$, and $R$ values. Like the case for TNG50 galaxies, the $R$ values in MaNGA are de-projected to be in the face-on orientation. Any spaxels containing bad data such as ill-defined stellar masses and/or undetected star formation rates are excluded from the comparison study. Additionally, to mimic the MaNGA detection thresholds in the simulation data, we limit our analysis to only the subset of all spaxels that obey $\Sigma_{\star} \ge 2 \times 10^7 {\rm M}_{\odot}$ kpc$^{-2}$ and $\Sigma_{\rm sfr} \ge 10^{-3} {\rm M}_{\odot}$ yr$^{-1}$ kpc$^{-2}$.

\subsection{Inference and Discussion}\label{sec6.3}

In Figure \ref{fig:fig9}, we present the results of our comparitive analysis between the observational and theoretical datasets. The top panels illustrate the individually normalised unweighted and $\Sigma_{\rm sfr}$-weighted distributions of log $\Sigma_{\star}$, log $\Sigma_{\rm sfr}$, and log $R$ as dotted and solid curves respectively. As the figure indicates, property distributions in MaNGA cover approximately the same range as TNG50 both in their original unweighted as well as the star formation-weighted forms. As a consequence, we also achieve an overall good agreement between the two in terms of peak height. Interestingly, due to the observational mocking steps involved, the TNG50 distributions no longer show a discernible bimodal shape, keeping in line with their observational counterparts. Specifically, applying the MaNGA point spread function on the simulation results reduces the $\Sigma_{\rm sfr}$ values of the inner pixels (at $R$ $\lesssim$ 2 kpc) by spreading star formation across multiple surrounding pixels. This in turn leads to an increase in the scale length of the otherwise steep inner exponential part of the TNG50 radial $\Sigma_{\rm sfr}$-profile and the smoothening of the elbow following it. This change thereby not only causes the suppression of the `inner' mode (c.f. \S\ref{sec4.5}), but also fades the prominent separation between the two modes by decreasing the difference between the slopes of the inner and outer exponential components of the profile.

Notwithstanding the general conformity, property distributions in TNG50 exhibit a few minor departures from that of MaNGA. The weighted MaNGA distributions of $\Sigma_{\star}$ and $\Sigma_{\rm sfr}$ seemingly favour peak values that are slightly greater in comparison with those exhibited by the TNG50 curves, a behaviour that is not perceptible in the case of their unweighted distributions. Another marginal disparity is apparent between the two $R$-distributions in that the TNG50 curve features a minute overall shift ($\approx$ 0.1 dex) towards higher values both in the unweighted and weighted forms relative to MaNGA. Lastly, the imposed hard $\Sigma_\star$-cut in the case of TNG50 expectedly manifests as a sharp lower limit of the unweighted distribution against a much softer, tapered edge in MaNGA. Curiously, while the peak of the weighted $R$ distribution in MaNGA approximately lines up with the outer (large-radius) mode in TNG, the peak in $\Sigma_{\star}$ is more compatible with the inner higher-density mode of the corresponding simulation-derived distribution.

The bottom panel of Figure \ref{fig:fig9} depicts the average $\Sigma_{\rm sfr}$ profiles for the two datasets, constructed in a similar fashion to the curve shown in Figure \ref{fig:fig5}. Barring a slight offset in the normalisations at large radii ($\approx$ 0.1 dex at $\ge$ 7 kpc), we find that the two profiles exhibit an excellent agreement with one another. In contrast with the 2-component profile composition discussed in \S\ref{sec4.5}, both the observed MaNGA and `mock' TNG50 profiles have a smooth shape that can be best described by a single continuous component. Indeed, if we fit the two curves, we find that the MaNGA(TNG) profile can be well-represented by a single exponential of scale length $\simeq 1.3(0.7)$ kpc plus a constant. We therefore deem the lack of two disparate scale lengths in this case to be of direct consequence to the shapes of the weighted distributions, explaining the absence of a bimodality. Finally, we notice that in spite of being largely akin to observations, the TNG-profile near the centre (at $R<1$ kpc) is somewhat shallower and has a value that is a factor of $\approx$ 2 lower compared to the inferred value in MaNGA. This difference could be partially responsible for the suppressed weighted probability i.e., star formation, at small-radii in TNG50 relative to MaNGA. Although, in light of our chosen simulation sampling scale and the spatial resolution of MaNGA ($\simeq$ 1 and 1.8 kpc respectively), we note that the nature of both profiles and their mutual comparison at these small radii may be inadequate to draw any robust conclusions. At large radii, the TNG50 curve has a steeper decline as indicated by the smaller best-fit scale length, while MaNGA retains a roughly constant value of star-formation albeit with a relatively noisier profile due to a significant dearth of star-forming spaxels in the outskirts.

Despite a careful galaxy sample selection and the application of MaNGA-like detection limits on simulation-derived data, the presence of minor dissimilarities between TNG50 and MaNGA suggests that the true extent of the selection effects inherent to the survey is perhaps not fully captured in the hard thresholds that we have used in our analysis. The resulting discrepancies are conspicuously reflected in the (difference between) unweighted distributions of $\Sigma_\star$ and $R$, as well as the $\Sigma_{\rm sfr} (R)$ profiles. As an example, the difference between the average $\Sigma_{\rm sfr}$-profile slopes at large radii and in central regions of galaxies is symptomatic of that fact that low-$\Sigma_{\rm sfr}$ regions are preferentially underrepresented in the observational data set that we have utilised. Such regions of low star formation are typically found in galaxy outskirts as well as in the centres of galaxies undergoing inside-out quenching (the latter would be expected due to our inclusion of cLIER galaxies). Considering that the properties of TNG galaxies such as sizes, star formation rates and specific star formation rates have previously been shown to be consistent with observed galaxy properties \citep[e.g.,][]{stevens19, hwang19, genel18}, we hypothesize that the dearth of these high-radius/central low-SFR pixels is borne out of vulnerability to the nuances of survey detection limits, particularly in $\Sigma_{\rm sfr}$. An exhaustive future study conducted using carefully generated mock-MaNGA observations from TNG will allow us to test this hypothesis. Nonetheless, we hope that with future resolved spectroscopic surveys pushing the detection limits, the trends emerging from our study could be directly tested against ISM property distributions in resolved observations, and provide the much needed clarity in this picture.

\section{Summary and outlook}\label{sec7}

In this work, we use the TNG50 cosmological simulation volume to generate and statistically survey the multi-dimensional parameter space of resolved ISM properties across a wide range of galaxy masses and redshifts. Specifically, we select star-forming galaxies in the mass range $10^7 - 10^{11}$ M$_\odot$ at $z$ = \{0, 0.5, 1, 2, 3\}. This sample accounts for more than 80\% of the total star formation in the simulation at each of the utilised snapshot. By dividing the galaxy into kpc-sized regions, we conduct a coarse-grained measurement of gas/stellar surface densities, gas metallicity, stellar vertical velocity dispersion, disc epicyclic frequency, star formation rate density and dark-matter volumetric density representative of each region (see \S\ref{sec3.2} for further details). We present a synopsis of the main findings of our analyses below.
\begin{itemize}[leftmargin=*]
    \item The distributions of all ISM properties, with the exception of stellar vertical velocity dispersion and gas surface density, exhibit bimodally-shaped distribution functions when weighted by star formation (Figure \ref{fig:fig2}), indicating that star formation in galaxies takes place in two separate environmental regimes. Star formation is most favoured to occur in the outer low-density low-metallicity regions for high-mass galaxies (above $\gtrsim 10^9 {\rm M}_\odot$), while being localised to the central high-density regions for lower mass galaxies (Figure \ref{fig:fig3}). For galaxies in a fixed mass bin, the preference for star formation in the outer diffuse regions is greater  at lower redshifts (Figure \ref{fig:fig4}). Additionally, our results show that most of the star formation in the universe takes place in galaxies with M$_{\star}$ = 10$^{10-11}$ M$_\odot$ at $z \leq 2$ (Figure \ref{fig:fig6}).
    
    \item The presence of a bimodality in property distributions results from an underlying bi-component average radial star formation rate profile for the galaxy sample. By fitting this profile with a combination of two exponentials, we demonstrate that the two peaks in the weighted distributions can be individually reproduced from the ``inner" and ``outer" exponential components separately (Figure \ref{fig:fig5}). We also find that almost all galaxies sustain a finite amount of star formation in both modes, albeit with varying degrees of relative contributions to them (Figure \ref{fig:fig8}, \ref{fig:fig8-2}).
    
    \item We investigate the 2D joint density distributions between parameters (Figure \ref{fig:fig10}), and find a very high degree of multi-collinearity (aka redundancy) in the 8D space. Through linear dimensionality-reduction via principal component analysis, we find that almost all of the intrinsic variance of the parameter space can be well captured via a transformed 3-dimensional representation (Figure \ref{fig:fig12}, \ref{fig:fig15}). Moreover, the leading principal component alone also captures the multi-parameter bimodality signature present in the original space (Figure \ref{fig:fig11}, \ref{fig:fig14}). This signature is manifested in the form of a ``radius-relation" (Figure \ref{fig:fig13}), i.e., the anti-correlation of galactocentric radius with the rest of the bimodally distributed ISM parameters.
    
    \item We conduct a preliminary comparison of our 1D property distributions and star formation rate profiles with those obtained from the MaNGA IFU survey for $z = 0$, with both galaxy samples selected to represent similar size-mass distributions (Figure \ref{fig:fig7}, \ref{fig:fig9}). Upon the application of observational detection limits, the $\Sigma_{\rm sfr}$-weighted distributions in TNG50 lose their bimodal shape showing concordance with the shape of the observed resolved property distributions. The comparison reveals an overall good match between TNG50 and MaNGA for the spread and peak locations of the parameter distributions, as well the the underlying average radial $\Sigma_{\rm sfr}$ profiles below $\simeq$ 7 kpc. We argue that some of the minor deviations between MaNGA and TNG50 results possibly arise from our inability to fully capture the nuances of the observational detection limits necessary to make a maximally unbiased comparison.
\end{itemize}

We envision that the results from our study will provide impetus for the construction of new heuristic star formation/ISM prescriptions in the near-future that are driven by fewer free parameters compared to currently used subgrid models. Given that our analysis is based on a dataset acquired from a fully cosmological, high-resolution, large volume simulation, we provide strong constraints for any model that endeavours to physically describe star formation and ISM physics in galaxies. By facilitating an optimal sampling of realistic initial conditions for future high-resolution ‘tall-box’ ISM simulations (e.g. TIGRESS), our characterisation will allow us to bridge the gap between stellar and galactic scales by establishing a direct link between small-scale ISM conditions and large-scale outflow properties. Finally, our work will provide avenues for meaningful comparison with similar measurements conducted with other large-scale cosmological simulations, as well as detailed quantitative patterns emerging from future high-precision spatially-resolved observations, especially at high-redshifts. \\

% B.M. is grateful to Francesco Belfiore for providing the MaNGA data catalogue used for the observational comparison presented in this paper, and for useful advice.
B.M. is grateful to Christopher C. Hayward and Tjitske Starkenburg for thoughtful suggestions that helped improve this manuscript. This work was carried out as part of the SMAUG project. SMAUG gratefully acknowledges support from the Center for Computational Astrophysics at the Flatiron Institute, which is supported by the Simons Foundation. The computational analyses presented in this paper were conducted in part on the Iron Cluster at the Flatiron Institute and the Popeye-Simons System at the San Diego Supercomputing Centre. B.M. thanks the scientific computing staff at the Flatiron Institute for their technical assistance. The work of C.-G.K. was partly supported by a grant from the Simons Foundation (CCA 528307, E.C.O.) and NASA ATP grant No. NNX17AG26G. E.C.O acknowledges support received under the Simons Investigator grant number 510940. The work of M.C.S. was supported by a grant from the Simons Foundation (CCA 668771, L.E.H.). TNG50 was run on the Hazel Hen supercomputer at the High Performance Computing Center Stuttgart using compute time granted by the Gauss Centre for Supercomputing (GCS) under GCS Large-Scale Project GCS-DWAR (CoPIs: Nelson, Pillepich).

\bibliography{refs}
\bibliographystyle{aasjournal}

\appendix
\section{Convergence with simulation resolution and pixel size} \label{appendixA}

\begin{figure}[h!]
\gridline{\fig{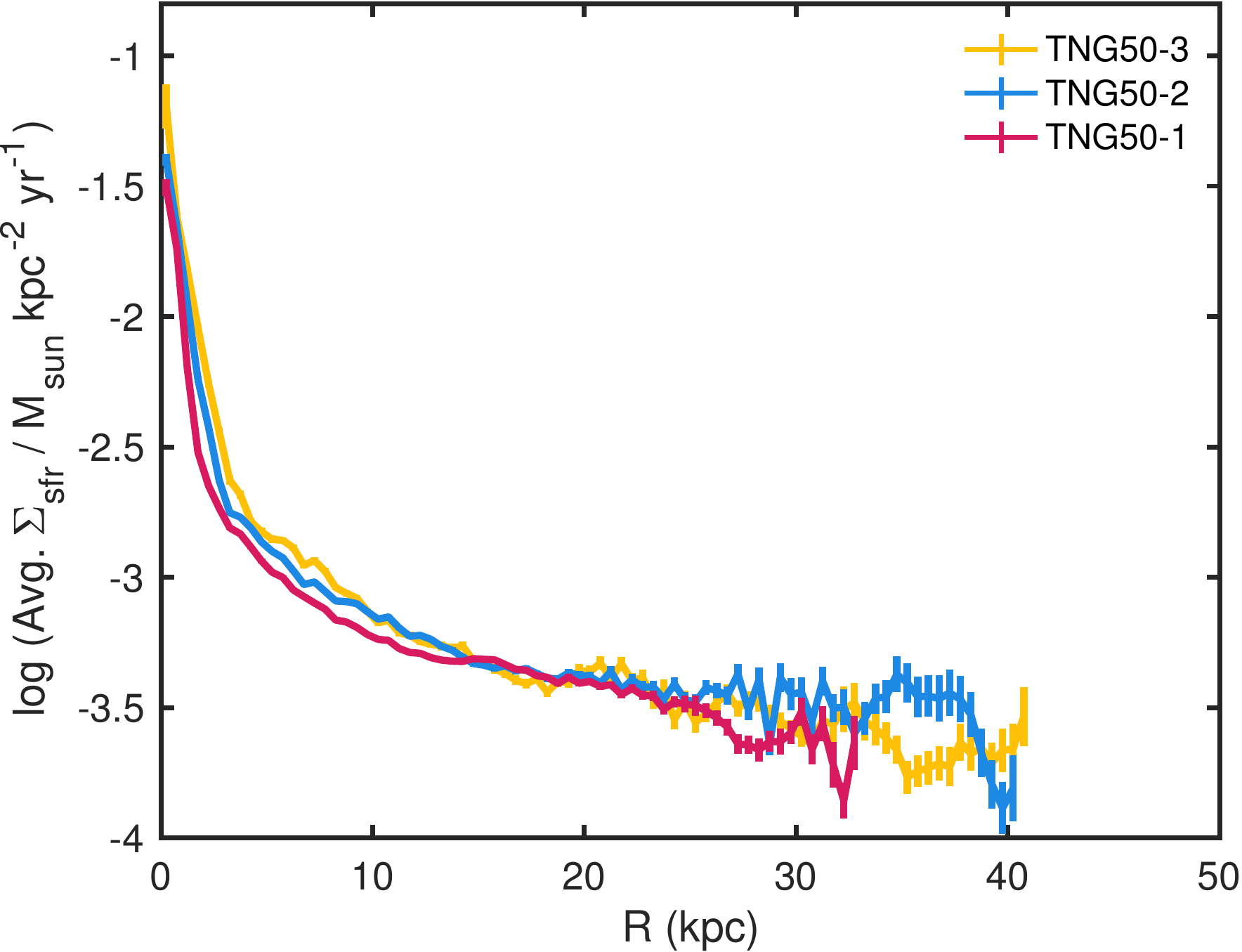}{0.48\textwidth}{}
          \fig{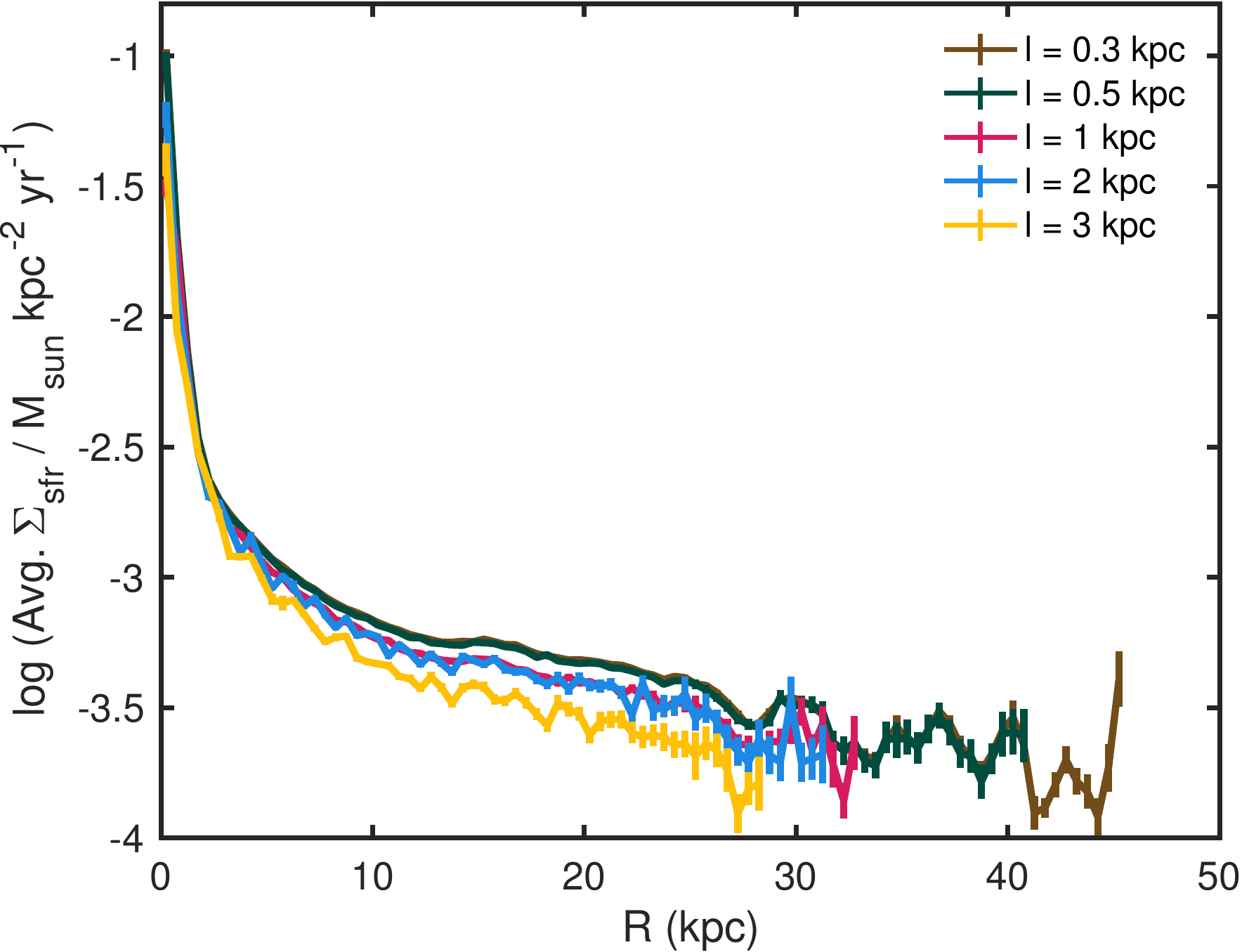}{0.48\textwidth}{}
         }
\caption{$\Sigma_{\rm sfr}$ profile variation with the simulation resolution (left) and choice of pixel size (right) for star-forming regions in our fiducial galaxy sample (M$_\star = 10^{9-10}$ at $z = 0$). In keeping with Figure \ref{fig:fig9}, only star-forming pixels are used to construct these profiles.}
\label{fig:fig16}
\end{figure}

We test the convergence of our results by way of comparing the average star formation rate profiles (shown in Fig. \ref{fig:fig9}) in Figure \ref{fig:fig16}. We use the two lower resolution, same volume counterparts of the standard TNG50-1 run, namely TNG50-2 and TNG50-3. TNG50-2(TNG50-3) has a mass-resolution of 6.8(54.2) M$_\odot$ for gas, and 36.3(290.4) M$_\odot$ for dark matter, making it roughly a factor of 8(64) coarser compared to TNG50-1. From the left panel in the figure, we observe that using a coarser-resolution box has marginal influence on our results both in terms of slope and normalisation, except at $R \geq 20$ kpc. Given the mass range we are working with, we expect these regions to correspond to the far-outskirts of the galaxies with very few star-forming gas particles, thus making the profile noisier due to poorer sampling.

In the right panel of the figure, we show a comparison of the same profile in TNG50-1, but this time for different choices of pixel size (image resolution). The profiles do not appear to have a dependence on pixel size for pixels in the central parts where the simulation resolution elements are smaller due to higher densities. In these regions, star formation is adequately resolved with a few 10s of particles contributing to each pixel on average. At large radii, we start to a see a weak variation of the slope with pixel size, such that bigger pixels give rise to steeper profiles. This is because in galaxy outskirts, where star formation does not have a uniform coverage, larger pixels tend to smooth over small-scale spatial patterns hence acquiring increasingly lower values of $\Sigma_{\rm sfr}$ as we go farther out. Lastly, as these profiles are binned and only composed of star-forming spaxels, they become noisier in the outskirts due to arbitrarily low area coverage as well as worsening Poissonian statistics.

\section{Reconstructing the original space from PCA results} \label{appendixB}
Let \textbf{X} be the data matrix corresponding to the initial space with \emph{n} rows for the samples and \emph{m} columns denoting features ($m = 8$ in our study). As described in \S\ref{sec5.2.1}, we standardise our dataset before conducting PCA by first subtracting the mean vector $\boldsymbol{\mu}$ from all rows and dividing them element wise by the standard deviation vector $\boldsymbol{\sigma}$. In our case, we made $\boldsymbol{\sigma}$ to be the $\Sigma_{\rm sfr}$-weighted standard deviations. Doing this gives us the corresponding matrix of standardised data, also known as z-scores \textbf{Z}.
After the PCA analysis, we obtain our results in the form of a coefficient matrix \textbf{C}, which is an $m \times m$ matrix whose columns are the $m$ eigenvectors representing the directions of the principal components (PCs). Then, the principal component scores are nothing but a projection of our original space along each of the PC directions. These are given in matrix form as \textbf{P} = \textbf{Z}\textbf{C}, with the same dimensionality as our original parameter space, i.e., $n \times m$.

Now, in order to reconstruct the original data from these scores, we apply the inverse operation such that $\boldsymbol{{\rm Z}} = \boldsymbol{\rm P}\boldsymbol{\rm C^{-1}}$. Due to the orthonormality of \textbf{C}, this is equivalent to \framebox{$\boldsymbol{{\rm Z}} = \boldsymbol{\rm P}\boldsymbol{\rm C^{T}}$}.
Finally, to obtain the reconstructed \textbf{X}, we multiply each column of \textbf{Z} by the corresponding $\sigma$ value and add the corresponding mean $\mu$. To obtain an approximate reconstruction using only a few PCs, say $k$ in number, one would only use the first $k$ PC scores, keeping just the first $k$ columns of \textbf{C} in the calculation above. 

\end{document}